\documentclass[10pt,a4paper,twoside]{article}

\pdfoutput=1

\usepackage{graphics}

\usepackage{amsmath}
\usepackage{sidecap}
\makeatletter
\def\SC@figure@vpos{t}   % align the caption with the top
\makeatother

\usepackage{multicol}
\usepackage[square,comma,numbers,sort&compress]{natbib}
\IfFileExists{microtype.sty}{\usepackage{microtype}}{\relax}

\usepackage{hyperref}

\hypersetup{  
pdftitle={Applications of quantum Monte Carlo methods in condensed systems},  
pdfauthor={Jindrich Kolorenc and Lubos Mitas},
pdfkeywords={quantum Monte Carlo, electronic structure calculations,
  computational methods, condensed matter theory, correlated
  electrons},
} 

\IfFileExists{glyphtounicode.tex}{%
\input glyphtounicode
\pdfgentounicode=1
}{\relax}

\textwidth=\paperwidth
\advance\textwidth -\oddsidemargin
\advance\textwidth -\evensidemargin
\newdimen\bottommargin
\textheight=\paperheight
\advance\textheight -\topmargin
\advance\textheight -\headheight
\advance\textheight -\headsep
\advance\textheight -\footskip
\advance\textheight -\bottommargin

\def\erfc{\mathop{\rm erfc}\nolimits}

\mathchardef\mhyphen="2D

\makeatletter

\DeclareMathAlphabet{\bi}{OML}{cmm}{b}{it}

\newcommand{\boldarrayrulewidth}{1\p@} 

\def\bhline{\noalign{\ifnum0=`}\fi\hrule \@height  
\boldarrayrulewidth \futurelet \@tempa\@xhline}

\def\@xhline{\ifx\@tempa\hline\vskip \doublerulesep\fi
      \ifnum0=`{\fi}}

\newcommand{\br}{\ms\bhline\ms}
\newcommand{\mr}{\ms\hline\ms}

\newcommand{\ms}{\noalign{\vspace{3\p@ plus2\p@ minus1\p@}}}
\newcommand{\bs}{\noalign{\vspace{6\p@ plus2\p@ minus2\p@}}}
\newcommand{\ns}{\noalign{\vspace{-3\p@ plus-1\p@ minus-1\p@}}}
\newcommand{\es}{\noalign{\vspace{6\p@ plus2\p@ minus2\p@}}\displaystyle}%

\def\0{\hbox{\phantom{\footnotesize\rm 0}}}%
\def\m{\hbox{$\phantom{-}$}}%
\newcommand{\rme}{\mathrm{e}}
\newcommand{\rmi}{\mathrm{i}}
\newcommand{\rmd}{\mathrm{d}}

\let\eref=\eqref
\let\fl=\relax

\newcommand{\Or}{\mathord{\mathrm{O}}}

\renewcommand\tableofcontents{\@starttoc{toc}}

\renewcommand\section{%
\@startsection {section}{1}{\z@}%
   {-3.5ex \@plus -1ex \@minus -.2ex}%
   {1.5ex \@plus.2ex}%
   {\normalfont\large\bfseries\raggedright}}
\renewcommand\subsection{%
\@startsection{subsection}{2}{\z@}%
   {-3.25ex\@plus -1ex \@minus -.2ex}%
   {1.5ex \@plus .2ex}%
   {\normalfont\normalsize\bfseries\raggedright}}
\renewcommand\subsubsection{%
\@startsection{subsubsection}{3}{\z@}%
   {-3.25ex\@plus -1ex \@minus -.2ex}%
   {1.5ex \@plus .2ex}%
   {\normalfont\normalsize\itshape\raggedright}}

\let\captionsize\small
\def\capfraction{1.0}
\newcommand{\mojec@ption}[2]{
\vskip\abovecaptionskip
\captionsize
\noindent\hbox to\textwidth{\hfill%
\vbox{\hsize=\capfraction\textwidth {\bfseries #1:} #2\hfill}%
\hfill}
\normalsize
\vskip\belowcaptionskip
}
\let\@makecaption\mojec@ption

\def\ps@simplefoot{
\def\@oddhead{\hfill}
\def\@evenhead{\hfill}
\def\@oddfoot{\hfill\thepage}
\def\@evenfoot{\thepage\hfill}
}
\makeatother

\begin{document}

\noindent
{\Large\bfseries Applications of quantum Monte Carlo methods in condensed
  systems}
\smallskip

\noindent
\begin{minipage}{.8\textwidth}

\noindent
{\large\sffamily Jind\v rich Koloren\v c}

\medskip
\hskip .03\textwidth
\begin{minipage}{.97\textwidth}
\begin{raggedright}
Institute of Physics, Academy of Sciences of the Czech Republic,
Na Slovance~2,\\ 18221 Praha~8, Czech Republic

\smallskip
I. Institut f\"ur Theoretische Physik, Universit\"at Hamburg,
  Jungiusstra\ss e~9,\\ 20355 Hamburg, Germany

\smallskip
{\tt kolorenc@fzu.cz}
\end{raggedright}
\end{minipage}

\bigskip

\noindent
{\large\sffamily Lubos Mitas}

\medskip
\hskip .03\textwidth
\begin{minipage}{.97\textwidth}
\begin{raggedright}
Department of Physics and Center for High Performance Simulation,
North Carolina State University, Raleigh, North Carolina 27695, USA

\smallskip
{\tt lmitas@unity.ncsu.edu}
\end{raggedright}
\end{minipage}

\bigskip
\noindent
The quantum Monte Carlo methods represent a powerful and broadly
applicable computational tool for finding very 
accurate solutions of the stationary Schr\"odinger equation for atoms,
molecules, solids and a variety of model systems. The algorithms are
intrinsically parallel and are able to take full advantage of the present-day
high-performance computing systems.
This review article concentrates on the fixed-node/fixed-phase
diffusion Monte Carlo method with emphasis on its applications
to electronic structure of solids and other extended many-particle
systems.

\end{minipage}
\bigskip

% 00.      GENERAL
% 02.      Mathematical methods in physics
% 02.70.-c Computational techniques; simulations
% 02.70.Ss Quantum Monte Carlo methods
%
% 70.      CONDENSED MATTER: ELECTRONIC STRUCTURE, ELECTRICAL,
%          MAGNETIC, AND OPTICAL PROPERTIES
% 71.      Electronic structure of bulk materials
% 71.15.-m Methods of electronic structure calculations
%
% 30.      ATOMIC AND MOLECULAR PHYSICS
% 31.      Electronic structure of atoms and molecules: theory
% 31.15.-p Calculations and mathematical techniques in atomic and
%          molecular physics
%
% 00.      GENERAL
% 05.      Statistical physics, thermodynamics, and nonlinear
%          dynamical systems
% 05.30.-d Quantum statistical mechanics
% 05.30.Fk Fermion systems and electron gas
%

\noindent
PACS numbers: 02.70.Ss, 71.15.$-$m, 31.15.$-$p

\bigskip
\noindent
To appear in {\itshape Rep. Prog. Phys.}

\bigskip

\hrule
\begin{multicols}{2}
{\footnotesize\tableofcontents}
\end{multicols}
\hrule

\begin{multicols}{2}
\section{Introduction}
\label{sec:intro}

\vskip -.25em
Many properties of condensed matter systems can be calculated
from solutions of the stationary Schr\"odinger equation describing
interacting ions and electrons. The grand challenge
of solving the Schr\"odinger equation has been around from the dawn of quantum mechanics
and remains at the forefront of the condensed matter physics today
and, undoubtedly, for many decades to come.   

 The task of solving the  
 Schr\"odinger equation for systems of electrons 
and ions, and predicting the quantities of interest such as cohesion and 
binding energies, electronic gaps, crystal structures, variety of magnetic phases or  
formation of quantum condensates is nothing short of formidable.
Paul Dirac recognized this state of affairs already in 1929: ``The general theory of quantum
mechanics is now almost complete~\ldots~The underlying physical laws necessary for the mathematical
theory of a large part of physics and the whole chemistry are thus completely known, and
the difficulty is only that the exact application of these laws leads to equations much
too complicated to be soluble.''\cite{dirac1929} Arguably, this is the most
fundamental approach to the physics of condensed matter: Applications
of the rigorous quantum laws to models that 
are as close to reality as currently feasible. 

The goal of finding accurate solutions for stationary quantum states
is hampered by a number of difficulties inherent to many-body quantum
systems:
\begin{enumerate}
\item Even moderately sized
model systems contain anywhere between tens to thousands of
quantum particles.  Moreover, we are often
interested in expectation values in the thermodynamic limit that
is usually reached by extrapolations from finite sizes. Such
procedures typically require detailed information about the scaling of the
quantities of interest with the system size.
\item Quantum
particles interact and the interactions affect the nature of quantum
states. In many cases, the influence is profound.
\item The solutions have to
conform to quantum symmetries such as the fermionic antisymmetry
linked to the Pauli exclusion principle. This is a fundamental
departure from classical systems and poses different challenges which
call for new analytical ideas and computational strategies.
\item For meaningful comparisons with experiments, the required accuracy is
exceedingly high, especially when comparing with precise data from spectroscopic and
low-temperature studies.
\end{enumerate}

In the past, the most successful approaches to address these challenges were based
mostly on reductionist ideas.
The problem is divided into the dominant effects, which are treated explicitly, 
and the rest, which is then dealt with by approximate methods based on variety
of analytical tools: perturbation expansions, mean-field methods,
approximate transformations to known solutions, and so on.
The reductionist approaches have been gradually developed into a high level of sophistication and
despite their limitations, they are still the most commonly used strategies in many-body physics. 
 
The progress in computer technology has opened a new avenue for
studies of quantum (and many other) problems and has enabled
researchers to obtain results beyond the scope of analytic many-body
theories. The performance of current large computers makes
computational investigations of many-body quantum systems viable,
allowing predictions that would be difficult or impossible to make
otherwise. The quantum Monte Carlo (QMC) methods described in this
review provide an interesting illustration of what is currently
possible and how much the computational methods can enrich and make
more precise our understanding of the quantum world.

Some of the ideas used in QMC methods go back to the times 
before the invention of electronic computers. Already in 1930s
Enrico Fermi noticed similarities between
the imaginary time Schr\"odinger equation and the laws governing stochastic processes in 
statistical mechanics. In addition, based on memories of his collaborator 
Emilio Segr\`e, Fermi also envisioned stochastic methodologies for solving
the Schr\"odinger equation, which were very similar to concepts developed decades later.
These Fermi's ideas were acknowledged by Metropolis and
Ulam in a paper from 1949~\cite{metropolis1949}, where they outlined a stochastic approach
to solving various physical problems and discussed merits of
``modern'' computers for its implementation.
In fact, this group of scientists 
 at the Los Alamos National Laboratory
attempted to calculate the hydrogen molecule by a simple version of 
QMC in the early 1950s, around the same time when a pioneering work on the first  
Monte Carlo study of classical systems was published by Metropolis and
coworkers \cite{metropolis1953}.
In the late 1950s, Kalos initiated development of
QMC simulations and methodologies 
for few-particle systems and laid down the statistical and mathematical
foundations of the Green's function Monte Carlo method
\cite{kalos1962}. Eventually, simulations of large
many-particle systems became practicable as well. First came studies of
bosonic fluids modelling ${}^4$He
\cite{mcmillan1965,kalos1974,whitlock1979}, and later followed
investigations of extended fermionic systems exemplified by liquid ${}^3$He
\cite{ceperley1977,lee1981} and by the homogeneous electron gas
\cite{ceperley1978,ceperley1980}. Besides these applications to
condensed matter, essentially the same methods were in mid-seventies
introduced in quantum chemistry to study
small molecular systems \cite{anderson1975,anderson1976}.  
To date, various QMC methods were developed and applied to the
electronic structure of atoms, molecules and solids, to quantum
lattice models, as well as to nuclear and other systems with contributions from
many scientists.

The term ``quantum Monte Carlo'' covers several related stochastic 
methodologies adapted to determine ground-state, excited-state 
or finite-temperature equilibrium properties of a variety of quantum systems.
The word ``quantum'' is important 
since QMC approaches differ
significantly from Monte Carlo methods for classical systems. For
an overview of the latter see for instance \cite{landau_binder_book}.
QMC is not only a computational tool for large-scale problems, but it also
encompasses a substantial amount of
analytical work needed to make such calculations feasible.
QMC simulations often utilize results of the more traditional electronic structure
methods in order to increase efficiency of the calculations. These ingredients are 
combined to optimally balance the computational cost with achieved
accuracy.
The key point for gaining new insights 
is an appropriate analysis of the quantum states and associated many-body effects.
It is typically approached
iteratively: Simulations indicate the gaps in understanding of the
physics, closing these gaps is subsequently attempted and the
improvements are assessed in the next round. Such a process involves
construction of zero- or first-order approximations for the desired
quantum states, incorporation of new analytical insights, and
development of new numerical algorithms.

QMC methods inherently incorporate several types of internal checks, and
many of the algorithms used possess various rigorous bounds, such as
the variational property of the total energy.
Nevertheless, the coding and numerical aspects of the simulations are
not entirely error-proof and
the obtained results should be verified independently. Indeed, it is a part of the modern 
computational-science practice that several groups revisit the same
problem with independent software packages and confirm or challenge the results.
``Biodiversity'' of the available QMC codes on the scientific market
(including QWalk \cite{wagner2009}, QMCPACK \cite{qmcpack_ref}, CHAMP
\cite{champ_ref}, CASINO \cite{casino_ref}, QMcBeaver
\cite{qmcbeaver_ref} and others) provides 
the important alternatives to verify 
the algorithms and their implementations. This is clearly a rather labourious, 
slow and tedious process, nevertheless, 
experience shows that independently calculated results and predictions eventually
reach a consensus and such verified data become widely used standards.

In this overview we present QMC methods that solve the stationary
Schr\"odinger equation for condensed systems of interacting fermions in continuous space.
Conceptually very straightforward is the variational Monte Carlo (VMC) method,
 which builds on explicit construction
of {\em trial (variational) wave functions} using stochastic integration and 
parameter optimization techniques.
More advanced approaches represented by the diffusion Monte Carlo
(DMC) method are based on projection operators that find the ground state within a given symmetry class.
Practical versions of the DMC method for a large number of particles require dealing with
the well-known fermion sign problem originating in the antisymmetry of
the fermionic wave functions. The most commonly used approach to overcome this 
fundamental obstacle is the fixed-node approximation.
This approximation introduces the so-called
{\em fixed-node error}, which appears to be the key limiting factor in further increase in accuracy. 
As we will see in section~\ref{sec:applications}, the fixed-node error
is typically rather small and does not hinder calculation of
robust quantities such as cohesion, electronic gaps, optical excitations, 
defect energies or potential barriers between structural conformations. By robust we mean quantities 
which are of the order
of tenths of an electronvolt to several electronvolts. Nevertheless, the fixed-node errors can bias
 results for more subtle phenomena, such as magnetic ordering or effects related
to superconductivity. Development of strategies to alleviate
such biases is an active area of research.

Fixed-node DMC simulations are computationally rather demanding  
 when compared to the mainstream electronic structure
methods that rely on mean-field treatment of electron-electron interactions.
On the other hand, QMC calculations can provide unique insights into the nature of quantum
phenomena and can 
verify many theoretical ideas. As such, they can 
produce not only accurate numbers but also new understanding.
Indeed, QMC methodology is very much an example of ``it from bit'' paradigm,
alongside, for example, the substantial computational efforts in quantum chromodynamics,
which not only predict hadron masses but also  
 contribute to the validation of the fundamental theory \cite{davies2004etal,durr2008}. 
Just a few decades ago it was difficult to imagine that one would be able to solve
the Schr\"odinger equation for hundreds of electrons by means of an
explicit construction of the many-body wave function.
Today, such calculations are feasible using available computational resources. At the same time,  
there remains more to be done to make the methods more insightful, more efficient, and
their application less labourious. We hope that this review will contribute
to the growing interest in this rapidly developing field of research.

The review is organized as follows: The remainder of this section
provides mostly definitions and notations.  Section~\ref{sec:methods}
follows with description of the VMC and DMC methods.  The strategies
for calculation of quantities in the thermodynamic limit are presented
in section~\ref{sec:pbc}.  Section~\ref{sec:wf} introduces currently
used forms of the trial wave functions and their recently developed
generalizations. The overview of applications presented in
section~\ref{sec:applications} is focused on QMC calculations of
a variety of solids and related topics.

\subsection{Many-body stationary Schr\"odinger equation}
Let us consider a system of quantum particles such as electrons and ions 
interacting via Coulomb potentials. Since the masses of nuclei and electrons 
differ by three orders of magnitude or more,
the problem can be simplified with the aid of
the {\em Born--Oppenheimer approximation}, which separates the
electronic degrees of freedom from the slowly moving system of ions.
The electronic part of the non-relativistic Born--Oppenheimer hamiltonian 
  is given by
\begin{equation}\label{eq:BOhamiltonian}
\hat H=-\frac12 \sum_i \nabla_i^2
 -\sum_{i,I}\frac{Z_{I}}{|\bi{r}_i-\bi{x}_{I}|}
 +\sum_{j<i}\frac1{|\bi{r}_i-\bi{r}_j|}\,,
\end{equation}
where $i$ and $j$ label the electrons and $I$
runs over the ions with charges $Z_{I}$.
Throughout the review we employ the atomic units,
$m_{\rm e}=\hbar=e=4\pi\varepsilon_0=1$, where $m_{\rm e}$ is the
electron mass, $-e$ is the electron charge and $\varepsilon_0$ is the
permittivity of a vacuum.
We are interested in eigenstates $|\Psi_n\rangle$ of the stationary
 Schr\"odinger equation 
\begin{equation}\label{eq:stationary_schrodinger}
 {\hat H} |\Psi_n\rangle= E_n|\Psi_n\rangle\,.
\end{equation}
Colloquially, we call such solutions (either exact or approximate) and derived properties  
collectively the electronic structure.

An important step forward in calculations of the eigenstates 
was made by Hartree~\cite{hartree1928} and Fock~\cite{fock1930} by
establishing the simplest antisymmetric 
wave functions and by formulating the Hartree--Fock
(HF) theory, which correctly takes into account
the Pauli {\em exclusion principle}~\cite{pauli1925,pauli1927b}. The HF theory replaces the hard problem of many
interacting electrons with a system of independent particles in an effective,
self-consistent field. The theory was further developed by 
Slater~\cite{slater1930} and others, and it has become a starting point of many sophisticated
approaches to fermionic many-body problems.

For periodic systems, 
the effective free-electron theory 
and the {\em band theory} of Bloch~\cite{bloch1929}
were the first crucial steps towards our present understanding of the
real crystals. In 1930s,
Wigner and Seitz~\cite{wigner1933,wigner1934} performed the first quantitative
calculations of the electronic states in sodium metal. 
Building upon the homogeneous electron gas model, the
density-functional theory (DFT) was invented by Hohenberg and
Kohn~\cite{hohenberg1964} and further developed by Kohn and
Sham~\cite{kohn1965} who formulated the local density approximation
(LDA) for the exchange-correlation functional. These ideas were
later elaborated by including spin polarization~\cite{vonbarth1972},
by constructing the generalized gradient
approximation (GGA) \cite{langreth1983,perdew1996}, and by designing a
variety of orbital-dependent exchange-correlation functionals
\cite{perdew1981,anisimov1991,kummel2008}. The DFT has proved to be
very successful and has become the mainstream 
computational method for many applications, which cover not only
solids but also molecules and even nuclear and other systems
\cite{kurth1999,sousa2007}. The
density-functional theory together with the Hartree--Fock and
post-Hartree--Fock methods \cite{szabo1989} are relevant for our
discussion of quantum 
Monte Carlo methodology, since the latter uses the results of these
approaches as a reference and also for construction of the many-body
wave functions. Familiarity with the basic concepts of the
Hartree--Fock and density-functional theories is likely to make the
subsequent sections easier to follow, but we believe that it is not a
necessary prerequisite for understanding our exposition of the QMC
methods and their foundations.

\section{Methods}
\label{sec:methods}

\subsection{Variational Monte Carlo}
\label{sec:vmc}

In the variational Monte Carlo method, the ground state of a
hamiltonian $\hat H$ is approximated by some trial wave function
$|\Psi_{\rm T}\rangle$, whose form is chosen following a prior analysis of the
physical system being investigated. Functional forms
relevant to solid-state applications will be discussed later in
section~\ref{sec:wf}. Typically
 a number of parameters are introduced
into $|\Psi_{\rm T}\rangle$, and these parameters are varied to minimize
the expectation value
$E_{\Psi_{\rm T}^2}=\langle\Psi_{\rm T}|\hat H|\Psi_{\rm T}\rangle/
\langle\Psi_{\rm T}|\Psi_{\rm T}\rangle$ in order to bring the trial
wave function as close as possible to the actual ground state
$|\Psi_0\rangle$.

Wave functions of interacting systems are non-separable, and
the integration needed to evaluate $E_{\Psi_{\rm T}^2}$ is therefore a
difficult task. Although it is possible to write these wave functions
as linear combinations of separable terms, this tactic is
viable only for a limited number of particles, since the length of
such expansions grows very quickly as the system size increases. The
variational Monte Carlo method employs a stochastic integration
that can treat the non-separable wave functions directly. The
expectation value $E_{\Psi_{\rm T}^2}$ is written as
\begin{multline}
\label{eq:vmc_integral}
E_{\Psi_{\rm T}^2}
%=\frac{\langle\Psi_{\rm T}|\hat H|\Psi_{\rm T}\rangle}{%
%  \langle\Psi_{\rm T}|\Psi_{\rm T}\rangle}
=\int\frac{|\Psi_{\rm T}(\mathcal{R})|^2}{%
  \langle\Psi_{\rm T}|\Psi_{\rm T}\rangle}\,
  \frac{\bigl[\hat H\Psi_{\rm T}\bigr](\mathcal{R})}{%
  \Psi_{\rm T}(\mathcal{R})}\,\rmd^{3N}\mathcal{R} \\
\approx E_{\rm VMC}
=\frac1{\mathcal{N}} \sum_{i=1}^{\mathcal{N}}
  \frac{\bigl[\hat H\Psi_{\rm T}\bigr](\mathcal{R}_i)}{%
  \Psi_{\rm T}(\mathcal{R}_i)}\,,
% + \Or(\mathcal{N}^{-1/2}) \,,
%+\mathop{\rm Err}(\hat H,\Psi_{\rm T},\mathcal{N})
\end{multline}
where $\mathcal{R}=(\bi{r}_1,\bi{r}_2,\dots,\bi{r}_N)$ is a
$3N$-dimensional vector encompassing the coordinates of all $N$ particles
in the system and the sum runs over
$\mathcal{N}$ such vectors $\{\mathcal{R}_i\}$ sampled from the
multivariate probability density
$\rho(\mathcal{R})=|\Psi_{\rm T}(\mathcal{R})|^2/
\langle\Psi_{\rm T}|\Psi_{\rm T}\rangle$. The summand 
$E_{\rm L}(\mathcal{R})=\bigl[\hat H\Psi_{\rm T}\bigr](\mathcal{R})/
\Psi_{\rm T}(\mathcal{R})$ is usually referred to as the local
energy. We assume spin-independent hamiltonians, and therefore spin
variables do not explicitly enter the evaluation of the expectation
value~\eref{eq:vmc_integral}. This statement is further corroborated
in section~\ref{sec:wf_elementary} where the elementary properties of
the trial wave functions~$|\Psi_{\rm T}\rangle$ are discussed.

Equation \eref{eq:vmc_integral} transforms the multidimensional
integration into a problem of sampling a complicated probability
distribution. The samples $\{\mathcal{R}_i\}$ can be obtained such
that they constitute a Markov chain with transitions
$\mathcal{R}_{i+1}\leftarrow\mathcal{R}_i$ governed by a stochastic
matrix $M(\mathcal{R}_{i+1}\leftarrow \mathcal{R}_i)$ whose stationary
distribution coincides with the desired probability density
$\rho(\mathcal{R})$,
\begin{equation}
\label{eq:markov_stationary}
\rho(\mathcal{R}') =
\int M(\mathcal{R}'\leftarrow \mathcal{R})\rho(\mathcal{R})\,
\rmd^{3N}\mathcal{R} \quad\hbox{for all $\mathcal{R}'$.}
\end{equation}
After a period of equilibration, the members of the Markov sequence
sample the stationary distribution regardless of the starting point
of the chain, provided the matrix $M(\mathcal{R}'\leftarrow
\mathcal{R})$ is ergodic. Inspired by the way the samples
explore the configuration space, they are often referred to
as walkers.

The Markov chain can be conveniently
constructed with the aid of the Metropolis method
\cite{metropolis1953,hastings1970}. The transition matrix is
factorized into two parts,
$M(\mathcal{R}'\leftarrow \mathcal{R}_i)
=T(\mathcal{R}'\leftarrow \mathcal{R}_i)
A(\mathcal{R}'\leftarrow \mathcal{R}_i)$, which correspond to two
consecutive stochastic processes: A candidate $\mathcal{R}'$ for $(i+1)$-th
sample is proposed according to the probability
$T(\mathcal{R}'\leftarrow \mathcal{R}_i)$ and this move is then either
accepted with the probability $A(\mathcal{R}'\leftarrow
\mathcal{R}_i)$ or rejected with the probability $1-A(\mathcal{R}'\leftarrow
\mathcal{R}_i)$. If the move is accepted, the new member of
the sequence is $\mathcal{R}_{i+1}=\mathcal{R}'$, otherwise it is
$\mathcal{R}_{i+1}=\mathcal{R}_i$. The length of the chain is thus
incremented in either case. The acceptance probability
$A(\mathcal{R}'\leftarrow \mathcal{R}_i)$, complementing some given
$T(\mathcal{R}'\leftarrow \mathcal{R}_i)$ and $\rho(\mathcal{R})$ such
that the stationarity condition \eref{eq:markov_stationary} is
fulfilled, is not unique.
The choice corresponding to the Metropolis algorithm reads 
\begin{equation}
A(\mathcal{R}'\leftarrow \mathcal{R})
=\min\biggl[1,
 \frac{T(\mathcal{R}\leftarrow \mathcal{R}')\,\rho(\mathcal{R}')}{%
  T(\mathcal{R}'\leftarrow \mathcal{R})\,\rho(\mathcal{R})}\biggr]
\end{equation}
and depends only on
ratios of $T$ and $\rho$. Consequently, normalization of the trial
wave function $|\Psi_{\rm T}\rangle$ is completely irrelevant for the
Monte Carlo evaluation of the quantum-mechanical expectation
values. The freedom to choose the proposal probability 
$T(\mathcal{R}'\leftarrow \mathcal{R}_i)$ can be exploited to improve
ergodicity of the sampling, for instance, to make it easier to overcome
a barrier of low probability density $\rho$ separating two high-density
regions. A generic choice for $T(\mathcal{R}'\leftarrow
\mathcal{R}_i)$ is a Gaussian distribution centered at $\mathcal{R}_i$
with its width tuned to optimize the efficiency of the sampling.

The variational energy $E_{\rm VMC}$ is a stochastic variable, and an
appropriate 
characterization of the random error $E_{\rm VMC}-E_{\Psi_{\rm T}^2}$ is
thus an integral part of the variational Monte Carlo method. 
When the sampled local energies $E_{\rm L}(\mathcal{R}_i)$ are
sufficiently well behaved 
\cite{trail2008a}, this error can be represented by 
the variance of $E_{\rm VMC}$. In such
cases, the error scales as $\mathcal{N}^{-1/2}$ and is
proportional to fluctuations of the local energy. Reliable
estimation of the variance of $E_{\rm VMC}$ is a non-trivial affair
since the random samples $\{\mathcal{R}_i\}$ generated by means of the
Markov chain are correlated. These correlations are not 
known {\em a priori\/} and depend on the particular form of the transition
matrix $M$ that varies from case to case. Nevertheless, it is possible
to estimate the variance without detailed knowledge of the correlation
properties of the chain with the aid of the
so-called blocking method \cite{flyvbjerg1989}.

The fluctuations of the local energy $E_{\rm L}$ are reduced as the
trial wave function $|\Psi_{\rm T}\rangle$ approaches an eigenstate of the
hamiltonian, and $E_{\rm L}$ becomes a constant when $|\Psi_{\rm T}\rangle$ is
an eigenstate. In particular, it is crucial to remove as many
singularities from $E_{\rm L}$ as possible by a proper choice of the
trial function. Section~\ref{sec:wf_elementary} illustrates how it is
achieved in the case of the Coulomb potential that is singular at
particle coincidences.

The total energy is not the only quantity of interest and evaluation
of other ground-state expectation values is often desired. The
formalism sketched so far remains unchanged, only the local energy is
replaced by a local quantity
$A_{\rm L}(\mathcal{R})=\bigl[\hat A\Psi_{\rm T}\bigr](\mathcal{R})/
\Psi_{\rm T}(\mathcal{R})$ corresponding to a general operator $\hat
A$. An important difference between $A_{\rm L}$ and the local energy 
is that fluctuations of $A_{\rm L}$ do not vanish when $|\Psi_{\rm
  T}\rangle$ is an 
eigenstate of $\hat H$. These fluctuations can severely impact the
efficiency of the Monte Carlo integration in
$\langle\Psi_{\rm T}|\hat A|\Psi_{\rm T}\rangle/
\langle\Psi_{\rm T}|\Psi_{\rm T}\rangle$, and the random error
can decay even slower than $\mathcal{N}^{-1/2}$
\cite{trail2008a}. The trial wave function cannot be altered to
suppress the fluctuations in this case, but a modified operator $\hat
A'$ can often be constructed such that 
$\langle\Psi_{\rm T}|\hat A'|\Psi_{\rm T}\rangle=
\langle\Psi_{\rm T}|\hat A|\Psi_{\rm T}\rangle$ while the fluctuations
of $A_{\rm L}$ are substantially reduced
\cite{assaraf1999,assaraf2000,assaraf2007,toulouse2007b,badinski2010}.

\subsection{Diffusion Monte Carlo}
\label{sec:dmc}

The accuracy of the variational Monte Carlo method is limited by the
quality of the trial wave function $|\Psi_{\rm T}\rangle$. This limitation can
be overcome with the aid of the projector methods. In
particular, the diffusion Monte Carlo method
\cite{grimm1971,anderson1975,hammond1994,foulkes2001} employs an imaginary time
evolution 
\begin{equation}
\label{eq:dmc_projection}
|\Psi_{\rm D}(t)\rangle=\exp\Bigl(-\bigl[\hat H-E_{\rm T}(t)\bigr]t\Bigr)
 \,|\Psi_{\rm T}\rangle\,,
\end{equation}
where the energy offset $E_{\rm T}$ is introduced to maintain
the wave-function norm at a fixed value. Formal properties of
\eref{eq:dmc_projection} can be elucidated by expanding the trial
function $|\Psi_{\rm T}\rangle$ in terms of the 
hamiltonian eigenstates~\eref{eq:stationary_schrodinger}, which
readily yields
\begin{multline}
|\Psi_{\rm D}(t)\rangle=
 \exp\Bigl(-\bigl[E_0-E_{\rm T}(t)\bigr]t\Bigr)
 \biggl[|\Psi_0\rangle\langle\Psi_0|\Psi_T\rangle \\
  + \sum_{n=1}^{\infty}\rme^{-(E_n-E_0)t}|\Psi_n\rangle
    \langle\Psi_n|\Psi_T\rangle \biggr]\,.
\end{multline}
The ground state $|\Psi_0\rangle$ is indeed reached in the limit of large $t$ 
as long as the trial function was not orthogonal to $|\Psi_0\rangle$
from the beginning. The requirement of a finite norm of
$|\Psi_{\rm D}(t)\rangle$ translates into a formula
\begin{eqnarray}
E_0=\lim_{t\to\infty} E_{\rm T}(t)
\end{eqnarray}
that can be used to obtain the ground-state energy. An alternative
approach is to evaluate the matrix element
$E_{\Psi_{\rm D}\Psi_{\rm T}}=\langle\Psi_{\rm D}(t)|\hat H|\Psi_{\rm T}\rangle/ 
\langle\Psi_{\rm D}(t)|\Psi_{\rm T}\rangle$ that asymptotically
coincides with the ground-state energy, since $\langle\Psi_0|\hat
H|\Psi_T\rangle/\langle\Psi_0|\Psi_T\rangle=\langle\Psi_0|\hat H|\Psi_0\rangle/
\langle\Psi_0|\Psi_0\rangle$. The integration in
$E_{\Psi_{\rm D}\Psi_{\rm T}}$ can be performed stochastically in
analogy with the VMC method,
\begin{multline}
\label{eq:dmc_mixed_estimate}
E_{\Psi_{\rm D}\Psi_{\rm T}}
=\int\frac{\Psi_{\rm D}^*(\mathcal{R},t)\Psi_{\rm T}(\mathcal{R})}{%
  \langle\Psi_{\rm D}(t)|\Psi_{\rm T}\rangle}\,
  \frac{\bigl[\hat H\Psi_{\rm T}\bigr](\mathcal{R})}{%
  \Psi_{\rm T}(\mathcal{R})}\,\rmd^{3N}\mathcal{R} \\
\approx E_{\rm DMC}
=\frac{1}{\mathcal{N}}
\sum_{i=1}^{\mathcal{N}} E_{\rm L}(\mathcal{R}_i)\,,
\end{multline}
where the individual samples $\mathcal{R}_i$ are now drawn from a probability
distribution defined as $\rho(\mathcal{R},t)=
\Psi_{\rm D}^*(\mathcal{R},t)\Psi_{\rm T}(\mathcal{R})/
\langle\Psi_{\rm D}(t)|\Psi_{\rm T}\rangle$.

\subsubsection{Fixed-node/fixed-phase approximation}

The Monte Carlo integration indicated in \eref{eq:dmc_mixed_estimate}
is possible only if $\rho(\mathcal{R},t)$ is real-valued and
positive. Since the hamiltonians we usually deal with are symmetric
with respect to time reversal, the eigenfunctions can be chosen
real. Unfortunately, many-electron wave functions must necessarily
have alternating sign to comply with the fermionic antisymmetry. In
general, the initial guess $|\Psi_{\rm T}\rangle$ will have different
plus and minus
sign domains (also referred to as nodal pockets or nodal
cells) than the sought for ground-state wave function $|\Psi_0\rangle$, which
results in changing sign of $\rho(\mathcal{R},t)$. In certain special
cases, the correct sign structure of the ground state can be deduced
from symmetry considerations
\cite{bressanini2005a,bressanini2005b,bajdich2005}, but in a general
interacting system
 the exact position of the boundary between the positive and negative domains
(the so-called fermionic node) is unknown and is determined by the
quantum many-body physics \cite{klein1976}. A number of exact
properties of the fermionic nodes have been discovered
\cite{ceperley1991,mitas2006,mitas2006b,bajdich2009}, but a lot
remains to be done in order to transform this knowledge into
constructive algorithms for the trial wave functions.

The problem with the variable sign of $\rho(\mathcal{R},t)$ can be
circumvented by complementing the projection \eref{eq:dmc_projection}
with the so-called fixed-node constraint \cite{anderson1976},
\begin{equation}
\label{eq:fixed_node}
\Psi_{\rm D}(\mathcal{R},t)\Psi_{\rm T}(\mathcal{R}) \geq 0\quad
\hbox{for all $\mathcal{R}$ and all $t$}\,.
\end{equation}
Doing so, $\lim_{t\to\infty}|\Psi_{\rm D}(t)\rangle$ only approximates
$|\Psi_0\rangle$, since the 
projection cannot entirely reach the ground state if the
initial wave function $|\Psi_{\rm T}\rangle$ does not possess the exact
nodes. The total energy calculated with this fixed-node method
represents an upper-bound estimate of the true ground-state energy
because the projection \eref{eq:dmc_projection}
is restricted to a subspace of the whole Hilbert space when the 
constraint \eref{eq:fixed_node} is implemented
\cite{moskowitz1982,reynolds1982,foulkes1999}. The fixed-node
approximation has proved very fruitful in 
quantum chemistry \cite{manten2001,grossman2002} as well as for
investigation of the electronic 
structure of solids as testified by the applications reviewed in
section~\ref{sec:applications}.

In calculations of extended systems and especially metals, it is
beneficial to allow for boundary conditions that break the
time-reversal symmetry, since it facilitates reduction of finite-size
effects (section~\ref{sec:tabc}). The eigenfunctions are then
complex-valued and a generalization of the fixed-node
approximation is required. The constraint \eref{eq:fixed_node} is
replaced with
$\Psi_{\rm D}(t)=|\Psi_{\rm D}(t)|\,\rme^{\rmi\varphi_{\rm T}}\,,$
where $\varphi_{\rm T}$ is the phase of the trial wave function
$\Psi_{\rm T}=|\Psi_{\rm T}|\,\rme^{\rmi\varphi_{\rm T}}$
\cite{ortiz1993}. The phase~$\varphi_{\rm T}$ is held constant during
the DMC simulation to guarantee that $\rho(\mathcal{R},t)$ stays
non-negative for all $\mathcal{R}$
and~$t$. Additionally, a complex trial wave function~$|\Psi_{\rm T}\rangle$
causes the local energy $E_{\rm L}$ to be complex as well. The
appropriate modification of the estimate for the total
energy~\eref{eq:dmc_mixed_estimate} 
coinciding with the asymptotic value of~$E_{\rm T}(t)$ then reads
\begin{equation}
\label{eq:dmc_mixed_estimate_fixed_phase}
E_{\rm DMC}
=\frac{1}{\mathcal{N}}
\sum_{i=1}^{\mathcal{N}}
\mathop{\rm Re}\bigl[E_{\rm L}(\mathcal{R}_i)\bigr]\,.
\end{equation}
Analogous
 to the fixed-node approximation, the fixed-phase
method provides a variational upper-bound estimate of the true
ground-state energy. Moreover, the fixed-phase
approximation reduces to the fixed-node approximation when applied to
real-valued wave functions.

\subsubsection{Sampling the probability distribution}
\label{sec:dmc_sampling}

The unnormalized probability distribution that we wish to sample in
the fixed-phase DMC method,
\begin{equation}
\label{eq:mixed_distrib}
f(\mathcal{R},t)
=\Psi_{\rm D}^*(\mathcal{R},t)\Psi_{\rm T}(\mathcal{R})
=|\Psi_{\rm D}(\mathcal{R},t)||\Psi_{\rm T}(\mathcal{R})|\,,
\end{equation}
referred to as the mixed distribution, fulfills an equation of motion
\begin{align}
\label{eq:mixed_distr_eq}
&-\partial_t f(\mathcal{R},t)=
\nonumber \\
&\hskip 2.6em
   -\frac12\nabla^2 f(\mathcal{R},t)
   +\nabla\cdot\bigl[\bi{v}_{\rm D}(\mathcal{R})f(\mathcal{R},t)\bigr]
\nonumber \\
&\hskip 2.6em
   +f(\mathcal{R},t)\Bigl[\mathop{\rm Re}\bigl[E_{\rm L}(\mathcal{R})\bigr]
     -(1+t\,\partial_t) E_{\rm T}(t) \Bigr]
\end{align}
that is derived by differentiating
\eref{eq:dmc_projection} and \eref{eq:mixed_distrib} with respect to
time, combining the resulting expressions and rearranging the
terms. The drift velocity  $\bi{v}_{\rm D}$
introduced in \eref{eq:mixed_distr_eq} is defined as
$\bi{v}_{\rm D}=\nabla\ln|\Psi_{\rm T}|$ and $\nabla$
denotes the $3N$-dimensional gradient with respect to $\mathcal{R}$. 
The equation of motion is valid in this form only as long as the
kinetic energy is the sole non-local operator in the
hamiltonian. Strategies for inclusion of non-local pseudopotentials
will be discussed later in section~\ref{sec:pp}.
The case of the fixed-node approximation is
virtually identical to \eref{eq:mixed_distr_eq}, except that the local
energy is real by itself. The following discussion therefore applies
to both methods.

The time evolution of the mixed distribution $f(\mathcal{R},t)$
can be written in the form of a convolution
\begin{equation}
\label{eq:mixed_distrib_conv}
f(\mathcal{R},t)=\int G(\mathcal{R}\leftarrow\mathcal{R}',t)
f(\mathcal{R}',0) \,\rmd^{3N}\mathcal{R}'\,,
\end{equation}
where $f(\mathcal{R},0)=|\Psi_{\rm T}(\mathcal{R})|^2$ and the Green's
function $G(\mathcal{R}\leftarrow\mathcal{R}',t)=
\langle\mathcal{R}|\hat G(t)|\mathcal{R}'\rangle$ is a solution of
\eref{eq:mixed_distr_eq} with the initial condition
$G(\mathcal{R}\leftarrow\mathcal{R}',0)=\delta(\mathcal{R}-\mathcal{R}')$.
Making use of the Trotter--Suzuki formula
\cite{trotter1959,suzuki1985}, the Green's function is approximated by
a product of short-time expressions,
\begin{equation}
\label{eq:trotter}
\hat G(t)=\bigl[\underbrace{
\hat G_{\rm g/d}(\tau)\,
\hat G_{\rm diff}(\tau)\,
\hat G_{\rm drift}(\tau)}_{\displaystyle\hat G_{\rm st}(\tau)}
\bigr]^M+\Or(\tau)\,,
\end{equation}
where $\tau$ denotes $t/M$ and the exact solution of
\eref{eq:mixed_distr_eq} is 
approached as this time step goes to zero. Consequently, the DMC
simulations should be repeated for several sizes of the time step and an
extrapolation of the results to $\tau\to 0$
should be performed in the end. For simplicity, we show in
\eref{eq:trotter} only
the simplest Trotter--Suzuki decomposition which has a time step error
proportional 
to $\tau$. Commonly used are higher order approximations whose errors
scale as $\tau^2$ or $\tau^3$.
 The three new Green's
functions constituting the short-time approximation $\hat G_{\rm st}$
can be explicitly written as 
\begin{align}
\label{eq:G_drift}
&G_{\rm drift}(\mathcal{R}\leftarrow\mathcal{R}',\tau)
\\
&\hskip1.5em
 = \bigl[1-\tau\nabla\cdot\bi{v}_{\rm D}(\mathcal{R}')\bigr]
    \delta\bigl[\mathcal{R}-\mathcal{R}'
      -\bi{v}_{\rm D}(\mathcal{R}')\tau\bigr]
\nonumber \\
&\hskip1.5em
  \phantom{=}\,\,+\Or(\tau^2)\,,
\nonumber \\[.8em]
\label{eq:G_diff}
&G_{\rm diff}(\mathcal{R}\leftarrow\mathcal{R}',\tau)
\\
&\hskip1.5em
  =\frac1{(2\pi\tau)^{3N/2}}
    \exp\biggl[-\frac{(\mathcal{R}-\mathcal{R}')^2}{2\tau}\biggr]\,,
\nonumber \\[.8em]
\label{eq:G_w}
&G_{\rm g/d}(\mathcal{R}\leftarrow\mathcal{R}',\tau)
\\
&\hskip1.5em
 = \exp\Bigl[-\tau\Bigl(\mathop{\rm Re}\bigl[E_{\rm L}(\mathcal{R})\bigr]
    -E_{\rm T}(t) \Bigr)\Bigr]
    \delta\bigl[\mathcal{R}-\mathcal{R}'\bigr]\,,
\nonumber
\end{align}
and correspond to the three non-commuting operators
from the right-hand side of \eref{eq:mixed_distr_eq} in the order:
drift $\bigl(\nabla\cdot\bigl[\bi{v}_{\rm
  D}(\mathcal{R})\,\bullet\bigr]\bigr)$,
diffusion $\bigl(-\nabla^2/2\,\bullet\bigr)$ and growth/decay
$\bigl(\bullet\bigl[\mathop{\rm Re}[E_{\rm L}(\mathcal{R})]
-(1+t\,\partial_t)E_{\rm T}(t)\bigr]\bigr)$. The drift and diffusion Green's
functions preserve the normalization of $f(\mathcal{R},t)$ whereas the
growth/decay process does not.

%The simple factorization \eref{eq:trotter} can be replaced with
%higher-order formulae \cite{suzuki1985} that reduce the error term to a
%higher power of the time step $\tau$. Benefits of such an improvement
%are only partial, however, since the drift Green's function is known
%only in the short time asymptotics and the error $\Or(\tau^2)$
%indicated in \eref{eq:G_drift} would contribute as $\Or(\tau)$ in
%\eref{eq:trotter} anyway.

The factorization of the exact Green's function into the product of the
short-time terms forms the basis of the stochastic process
that represents the diffusion Monte Carlo algorithm. First,
$\mathcal{M}$ samples $\{\mathcal{R}_i\}$ are drawn from the distribution
$f(\mathcal{R},0)=|\Psi_{\rm T}(\mathcal{R})|^2$ just like in the VMC
method. Subsequently, this set of walkers evolves such that it samples
the mixed distribution $f(\mathcal{R},t)$ at any later time $t$.
The probability distribution is updated from time $t$ to $t+\tau$ by
multiplication with the short-time Green's function,
\begin{equation}
\label{eq:time_step_conv}
f(\mathcal{R},t+\tau)=\int G_{\rm st}(\mathcal{R}\leftarrow\mathcal{R}',\tau)
f(\mathcal{R}',t) \,\rmd^{3N}\mathcal{R}'\,,
\end{equation}
which translates into the following procedure performed on each walker in the
population:
\begin{enumerate}
\item \label{item:dmc_step1} a drift move
$\Delta\mathcal{R}_{\rm drift}=\bi{v}_{\rm D}(\mathcal{R}')\tau$ is
proposed
\item a diffusion move $\Delta\mathcal{R}_{\rm diff}=\boldsymbol{\chi}$
  is proposed, where $\boldsymbol{\chi}$ is a vector of Gaussian
  random numbers with variance $\tau$ and zero mean
\item the increment $\Delta\mathcal{R}_{\rm drift}
  +\Delta\mathcal{R}_{\rm diff}$ is accepted if it complies with the fixed-node
  condition $\Psi_{\rm T}(\mathcal{R}')
  \Psi_{\rm T}(\mathcal{R}'+\Delta\mathcal{R}_{\rm drift}
  +\Delta\mathcal{R}_{\rm diff})>0$, otherwise the walker stays at its
  original position;
  moves attempting to cross the node occur only due
  to inaccuracy of the approximate Green's function~\eref{eq:trotter},
  and they vanish in the limit $\tau\to 0$;
  the moves $\Delta\mathcal{R}_{\rm drift}+\Delta\mathcal{R}_{\rm
    diff}$ are accepted without any constraint in the fixed-phase
  method
\item \label{item:dmc_gw} the growth/decay Green's function $G_{\rm
    g/d}$ is applied; several algorithms devised for this purpose
  are outlined in the following paragraph
\item \label{item:dmc_sample_collect} at this moment, the time step is
  finished and the simulation continues with another cycle starting
  back at (\ref{item:dmc_step1}).
\end{enumerate}
After the projection period is completed, the algorithm
samples the desired ground-state mixed distribution and the quantities
needed for evaluation of various expectation values can be collected
in step~(\ref{item:dmc_sample_collect}).

At this point we return to a more detailed discussion of several algorithmic
representations of the growth/decay Green's function
$G_{\rm g/d}$ needed in step~(\ref{item:dmc_gw}).
\begin{itemize}
\item The most straightforward way is to assign a weight
  $w$ to each walker. These weights are set to 1
  during the VMC initialization of the walker population and the application of
  $G_{\rm g/d}$ then amounts to a multiplication
  $w(t+\tau)=w(t)W(\mathcal{R})$, where the weighting
  factor is
\begin{equation}
\label{eq:weight_factor}
W(\mathcal{R})=\exp\Bigl[
  -\tau\Bigl(\mathop{\rm Re}\bigl[E_{\rm L}(\mathcal{R})\bigr]
  -E_{\rm T}(t) \Bigr)\Bigr]\,.
\end{equation}
  Consequently, the formula for calculation of the total energy
  \eref{eq:dmc_mixed_estimate_fixed_phase} is modified to
\begin{equation}
\label{eq:dmc_mixed_estimate_pure_diff}
E_{\rm DMC}
=\biggl(\sum_{i=1}^{\mathcal{N}} w_i\biggr)^{-1}
\sum_{i=1}^{\mathcal{N}} w_i
\mathop{\rm Re}\bigl[E_{\rm L}(\mathcal{R}_i)\bigr]
\end{equation} 
  and the walkers remain distributed according to
  $|\Psi_{\rm T}(\mathcal{R})|^2$ as in the VMC method.
  This algorithm is referred to as the {\em pure diffusion Monte Carlo}
  method \cite{caffarel1988a,ceperley1988}. It is known to be
  intrinsically unstable at large projection times where the 
  signal-to-noise ratio deteriorates \cite{assaraf2000b}, but it
  is still useful for small quantum-chemical systems
  \cite{caffarel1988b,flad1997,schautz1999}.
\item The standard DMC algorithm fixes the weights to $w=1$ and
  instead allows for stochastically 
  fluctuating size of the walker population by branching
  walkers in regions with large weighting factor
  $W(\mathcal{R})$ and by removing them from areas with small
  $W(\mathcal{R})$.
%  The application of $G_{\rm w}$ is implemented
%  as a replacement of the original walker with its $n=\mathop{\rm int}
%  [W(\mathcal{R}''')+\eta]$ copies where $\mathop{\rm int}$ denotes
%  the integer part of a real number and $\eta$ is a random number
%  drawn from a uniform distribution on the interval $(0,1)$.
  The copies from high-probability regions are treated as
  independent samples in the subsequent time steps.
  The time dependence of the energy offset $E_{\rm T}(t)$ provides a
  population control mechanism that prevents the population from
  exploding or collapsing entirely \cite{umrigar1993,hammond1994}. The
  {\em branching/elimination} algorithm is much more efficient in large
  many-body systems than the pure DMC method, although it
  also eventually reaches the limits of its applicability for
  a very large number of particles \cite{nemec2010}. 
\item An alternative to the fluctuating population are
  various flavours of the {\em stochastic reconfiguration}
  \cite{hetherington1984,buonaura1998,assaraf2000b,jones2009,wagner2009}.
  These algorithms complement each branched
  walker with high weighting factor~$W(\mathcal{R})$ with
  one eliminated walker with small~$W(\mathcal{R})$, and therefore
  the total number of walkers remains constant. This pairing
  introduces slight correlations into the walker population
  that are comparable to those caused by the
  population control feedback of the standard branching/elimination algorithm
  \cite{nemec2010}. Keeping the population size fixed is
  advantageous for load balancing in parallel computational
  environments, since the number of walkers can be a multiple of
  the number of computer nodes (CPUs) at all times during the simulation.
\end{itemize}
The branching/elimination process interacts in a subtle way with the
fixed-node constraint. Since the walkers are not allowed to cross the node, the
branched and parent walkers always stay in the same nodal cell. If
some of these cells are more favoured (that is, if
they have a lower local energy on average), the walker population
accumulates there and eventually vanishes from the less favoured
cells. Such uneven distribution of samples would introduce
a bias to the simulation. Fortunately, it does not happen, since
all nodal cells of the ground-state wave functions are connected by
particle permutations and are therefore equivalent, see the tiling theorem in
\cite{ceperley1991}. For general excited states this theorem does not hold
and the unwanted depopulation of some nodal cells can indeed be
observed. The problem is absent from the fixed-phase
method, since it contains no restriction on the walker propagation.

The branching/elimination algorithm is just one of the options of
dealing with the weights along the stochastic paths. Another
possibility was 
introduced by Baroni and Moroni as the so-called {\em reptation}
algorithm \cite{baroni1999}, which recasts the sampling of both the
path in the configuration space {\em and} the weight into a
straightforward Monte Carlo process, avoiding thus some of the
disadvantages of the DMC algorithm. The reptation method
has its own sources of inefficiencies which can be, however,
significantly alleviated \cite{moroni_private}.

This concludes our presentation of the stochastic techniques
that are used to simulate the projection operator introduced in
\eref{eq:dmc_projection}. We would like to bring to the reader's
attention that the algorithm outlined in this section is rather
rudimentary and illustrates only the general ideas. A number of
important performance improvements are usually employed in practical
simulations, see for instance \cite{umrigar1993} for further details.

\subsubsection{General expectation values}

So far, only the total energy was discussed in connection with the DMC
method. An expression analogous to \eref{eq:dmc_mixed_estimate} can be
written with any operator $\hat A$ in place of the hamiltonian
$\hat H$. The acquired quantity
$A_{\Psi_{\rm D}\Psi_{\rm T}}=\langle\Psi_{\rm D}|\hat A|\Psi_{\rm T}\rangle/
\langle\Psi_{\rm D}|\Psi_{\rm T}\rangle$, called the mixed
estimate, differs from the pure expectation value $\langle\Psi_{\rm D}|\hat
A|\Psi_{\rm D}\rangle/\langle\Psi_{\rm D}|\Psi_{\rm D}\rangle$
 unless
$\hat A$ commutes with the hamiltonian. In general, the error is
proportional to 
the difference between $|\Psi_D\rangle$ and $|\Psi_{T}\rangle$. The
bias can be reduced to the next order using the following
extrapolation \cite{whitlock1979,hammond1994}
\begin{multline}\label{eq:extrap_estim}
\frac{\langle\Psi_D|\hat A|\Psi_D\rangle}{%
\langle\Psi_D|\Psi_D\rangle}
 = 2\, \frac{\langle\Psi_D|\hat A|\Psi_{T}\rangle}{%
    \langle\Psi_D|\Psi_{T}\rangle}
 - \frac{\langle\Psi_{T}|\hat A|\Psi_{T}\rangle}{%
    \langle\Psi_{T}|\Psi_{T}\rangle} \\
   +\Or\biggl(\biggl|
      \frac{\Psi_D}{\sqrt{\langle\Psi_D|\Psi_D\rangle}}
      -\frac{\Psi_{T}}{\sqrt{\langle\Psi_{T}|\Psi_{T}\rangle}}
   \biggr|^2\biggr)\,.
\end{multline}
Alternative methods that allow for a direct
evaluation of the pure expectation values have been developed, such as
the forward (or future) 
walking \cite{liu1974,barnett1991,hammond1994}, the reptation quantum Monte
Carlo \cite{baroni1999,pirleoni2005,yuen2009}, or the Hellman--Feynman
operator sampling \cite{gaudoin2007b,gaudoin2010}. Due to their certain
limitations, 
these techniques do not fully replace the extrapolation
\eref{eq:extrap_estim}---they are usable only
for local operators and the former two become computationally
inefficient in large systems.

The discussion of the random errors from the end of
section~\ref{sec:vmc} applies also to the 
diffusion Monte Carlo method, except that the serial correlations
among the data produced in the successive steps of the DMC simulations
are typically larger than the correlations in the VMC data. Therefore,
longer DMC runs are necessary to achieve equivalent suppression of the
stochastic uncertainties of the calculated expectation values.

\subsubsection{Spin degrees of freedom}
\label{sec:dmc_spins}

The DMC method as outlined above samples only the spatial part
of the wave function, and the spin degrees of
freedom remain fixed during the whole simulation. This
simplification follows from the assumption of
a spin-independent hamiltonian that implies freezing of spins during
the DMC projection \eref{eq:dmc_projection}. This is indeed the current state
of the DMC methodology as applied to electronic-structure problems: In
order to arrive at the correct spin state,
a number of spin-restricted calculations are performed and the
variational principle is employed to select the best ground state
candidate among them.

Fixing spin
variables is not possible for spin-dependent hamiltonians, such as for
those containing spin-orbital interactions, since they lead to a
non-trivial coupling of different spin configurations. In fact, spin-dependent
quantum Monte Carlo methods were developed for studies of 
nuclear matter. A variant of the Green's function Monte Carlo method
\cite{carlson1987,pudliner1997} treats the spin degrees of freedom
directly in their full state space. Since the number of spin
configurations grows exponentially with the number of particles,
this approach is limited to relatively small systems. More favourable
scaling with the systems size offers the auxiliary field diffusion
Monte Carlo method that samples the spin variables stochastically by
means of auxiliary fields introduced via the Hubbard--Stratonovich
transformation \cite{schmidt1999,sarsa2003}. Recently, a version of the 
auxiliary field DMC method was used to investigate properties of the
two-dimensional electron gas in presence of the Rashba spin-orbital
coupling \cite{ambrosetti2009}.

\subsection{Pseudopotentials}
\label{sec:pp}

The computational cost of all-electron QMC calculations grows very rapidly
with the atomic number~$Z$ of the elements constituting the simulated
system. Theoretical analysis \cite{ceperley1986,hammond1987} as well
as practical calculations~\cite{ma2005b} indicate that the
cost scales as $Z^{5.5-6.5}$. Most of the computer time spent
in these simulations is used for sampling of large energy fluctuations
in the core region, which have very little effect on typical properties
of interest, such as interatomic bonding and low-energy
excitations. For investigations of these quantities it is convenient
to replace the core electrons with accurate pseudopotentials.
A sizeable library of norm-conserving
pseudopotentials targeted specifically to applications of the QMC methods
has been built over the years
\cite{lee2000,ovcharenko2001,trail2004,trail2005,burkatzki2007,burkatzki2008}.

Pseudopotentials substitute the ionic Coulomb potential with a
modified expression,
\begin{equation}
\label{eq:coulomb_to_pp}
-\frac{Z}{r}\quad\to\quad
  V(r)+\hat W\,
\end{equation}
where $V(r)$ is a local term behaving asymptotically as $-(Z-Z_{\rm
  core})/r$ with $Z_{\rm core}$ being the number of eliminated core
electrons. The operator $\hat W$ is non-local in the sense that
electrons with different angular momenta experience different radial
potentials. Explicitly, the matrix elements of the potential $\hat W$
associated with $I$-th atom in the system are
\begin{multline}
\label{eq:pp_mtrx_elem}
\langle\mathcal{R}|\hat W_I|\mathcal{R}'\rangle=
 \sum_i\sum_{l=0}^{l_{\rm max}}\sum_{m=-l}^l
   \langle\hat{\bi{r}}_{iI}|lm\rangle
\\
\times W_{I,l}(r_{iI})\delta(r_{iI}-r_{iI}')
   \langle lm|\hat{\bi{r}}_{iI}'\rangle\,,
\end{multline}
where $|lm\rangle$ are angular momentum eigenstates, $r_{iI}$ is the
distance of an electron from the $I$-th nucleus and $\hat{\bi{r}}_{iI}$
is the associated direction $\bi{r}_{iI}/r_{iI}$. Functions~$W_{I,l}$
vanish for 
distances $r_{iI}$ larger than some cut-off radius~$r_{\rm c}$, and
the sum $\sum_i$ therefore runs only over electrons that are
sufficiently close to the particular nucleus.

Evaluation of pseudopotentials in the VMC method is straightforward,
despite the fact that the local energy~$E_{\rm L}$ itself involves
integrals. As can be
inferred from the form of the matrix elements \eref{eq:pp_mtrx_elem},
these are two-dimensional integrals over surfaces of spheres centered
at the nuclei. The integration can be implemented with the aid of the
Gaussian quadrature rules that employ favourably sparse meshes
\cite{fahy1990,mitas1991}.

The use of non-local pseudopotentials in the fixed-node DMC method is more
involved, since the sampling algorithm outlined in
section~\ref{sec:dmc_sampling} explicitly assumed that all potentials
were local. Non-local hamiltonian terms can be formally
incorporated by introducing an extra member into the Trotter
break-up~\eref{eq:trotter}, namely
\begin{align}
\label{eq:gf_nloc_moves}
&G_{\rm nloc}(\mathcal{R}\leftarrow\mathcal{R'},\tau)
\nonumber\\
&\hskip .5em
= \frac{\Psi_{\rm T}(\mathcal{R})}{\Psi_{\rm T}(\mathcal{R}')}\,
 \langle\mathcal{R}|\rme^{-\tau\hat W}|\mathcal{R}'\rangle
\nonumber\\
&\hskip .5em
=\delta(\mathcal{R}-\mathcal{R}')
-\frac{\Psi_{\rm T}(\mathcal{R})}{\Psi_{\rm T}(\mathcal{R}')}\,
 \langle\mathcal{R}|\tau\hat W|\mathcal{R}'\rangle
+\Or(\tau^2)\,,
\end{align}
where $\hat W$ now combines the non-local
contributions from all atoms in the system. This
alone is not the desired solution, since the term involving the
matrix element of $\hat W$ does not have a fixed sign and thus cannot be
interpreted as a transition probability.

To circumvent this difficulty the so-called localization approximation
has been proposed. It amounts to a replacement of the non-local
operator in the hamiltonian with a local expression
\cite{hurley1987,hammond1987,mitas1991}
\begin{equation}
\hat W \quad\to\quad
W_{\rm L}(\mathcal{R})=
\frac{\hat W\Psi_{\rm T}(\mathcal{R})}{\Psi_{\rm T}(\mathcal{R})}\,.
\end{equation}
Consequently, the contributions from $\hat W$ are directly
incorporated into the growth/decay Green's function~\eref{eq:G_w} and
no complications with alternating sign arise.
Unfortunately, the DMC method with this approximation does not 
necessarily provide
an upper-bound estimate for the ground-state energy. The calculated
total energy
$E_{\rm DMC}$ is above the lowest eigenvalue of the localized
hamiltonian, which does not guarantee that it is
also higher than the ground-state energy of the original
hamiltonian~$\hat H$. The errors in the total energy incurred by the
localization approximation are quadratic in the difference between the
trial function~$|\Psi_{\rm T}\rangle$ and the exact ground-state wave
function \cite{mitas1991}. The trial wave
functions are usually accurate enough for the localization error to be
practically insignificant and nearly all applications listed in
section~\ref{sec:applications} utilize this approximation.

A method that preserves the upper-bound property of
$E_{\rm DMC}$ was proposed in the context of the DMC algorithm developed for
lattice models \cite{tenhaaf1995}. The non-local operator $\hat W$ is
split into two parts, $\hat W=\hat W^++\hat W^-$, such that $\hat W^+$
contains those matrix elements, for which $\langle\mathcal{R}|\hat
W|\mathcal{R}'\rangle\Psi_{\rm T}(\mathcal{R})\Psi_{\rm
  T}(\mathcal{R}')$ is positive, and $\hat W^-$ contains the elements, for
which the expression is negative. Only the $\hat W^+$ part is
localized in order to obtain the approximate hamiltonian,
\begin{equation}
\hat W \quad\to\quad
\hat W^- +
\frac{\hat W^+\Psi_{\rm T}(\mathcal{R})}{\Psi_{\rm T}(\mathcal{R})}\,.
\end{equation}
One can explicitly show that the lowest eigenvalue of
this partially localized hamiltonian is an upper
bound to the lowest eigenvalue of the original fully non-local
hamiltonian \cite{tenhaaf1995}.
Recently, a stochastic representation of the non-local Green's
function \eref{eq:gf_nloc_moves} corresponding to $\hat W^-$ was
implemented also into the DMC method for continuous space
\cite{casula2006}. Apart from the recovered upper-bound property, the
new algorithm reduces fluctuations of the local energy for certain
types of
pseudopotentials. On the other hand, the time step error is in general
larger \cite{casula2006,pozzo2008a}, since the distinct treatment of the
$\hat W^+$ and $\hat W^-$ parts of the pseudopotential essentially
corresponds to a Trotter splitting of the growth/decay Green's
function~\eref{eq:G_w} into two pieces. Very recently, a more accurate
Trotter break-up and other modifications improving efficiency
of this method have been proposed for both continuous and lattice DMC
formulations \cite{casula2010}.

The localization approximation is directly applicable also to the
fixed-phase DMC method. Adaptation of the non-local moves representing
$\hat W^-$ to cases
involving complex wave functions has not been reported yet,
nevertheless, the modifications required should be only minor.

\section{From a finite supercell to the thermodynamic limit}
\label{sec:pbc}

Quantum Monte Carlo methods introduced in the preceding chapter can be
straightforwardly applied to physical systems of a finite size, such as
atoms and clusters of atoms. To allow investigation of bulk
properties of solids, the algorithms described so far have to be
complemented with additional techniques that reduce the essentially
infinitely many degrees of freedom into a problem of manageable
proportions.

\subsection{Twist-averaged boundary conditions}
\label{sec:tabc}

In approximations
that model electrons in solids as an ensemble of independent
(quasi\nobreakdash-)particles, it is possible to map the whole infinite crystal
onto a finite volume so that the thermodynamic limit becomes directly
accessible. Hamiltonians of such models are invariant with respect to
separate translations of electrons by any lattice vector $\bi{R}$,
that is, for each $i$ we can write
\begin{multline}\label{eq:ham_transl_lda}
\hat H(\bi{r}_1,\bi{r}_2,\dots,\bi{r}_i+\bi{R},\dots) \\ =
 \hat H(\bi{r}_1,\bi{r}_2,\dots,\bi{r}_i,\dots)\,.
\end{multline}
This invariance allows to diagonalize the hamiltonian only in the primitive
cell of the lattice and then use the translations to expand
the eigenstates from there into the whole crystal. Unfortunately, the
explicit two-body interactions that we are 
set out to keep in the hamiltonian break the symmetry
\eref{eq:ham_transl_lda}. The only
translation left is a simultaneous displacement of all electrons by a
lattice vector,
which is not enough to reach the thermodynamic limit with a finite set
of degrees of freedom.

To proceed further we introduce artificial
translational symmetries with the aid of the so-called supercell
approximation that is widely used within the independent-particle
methods to investigate non-periodic structures such as lattice
defects. We select a supercell having a volume $\Omega_S$ that
contains several (preferably many) primitive cells. The whole
crystal is then reconstructed via translations of this large cell by
supercell lattice vectors $\{\bi{R}_S\}$, which are a subset of all
lattice vectors $\{\bi{R}\}$. Simultaneously, we divide the
electrons in the solid into groups containing $N=\rho_{\rm
  av}\Omega_S$ particles, where $\rho_{\rm av}$ is the average
electron density in the crystal. This partitioning is used to
construct a model hamiltonian, where electrons within each group
interact, whereas there are no interactions between the groups,
\begin{align}
\label{eq:supercell_ham_sum}
\hat H_{\rm model}&=\sum_{I=1}^{\infty}
  \hat H_S\bigl(\mathcal{R}^{(I)}\bigr)
\nonumber\\
&=\sum_{I=1}^{\infty}\Biggl[\sum_{i=1}^N \hat h\bigl(\bi{r}_i^{(I)}\bigr)
  +\hat V_{ee}\bigl(\mathcal{R}^{(I)}\bigr)\Biggr]\,.
\end{align}
The vector
$\mathcal{R}^{(I)}=\bigl(\bi{r}_1^{(I)},\dots,\bi{r}_{N}^{(I)}\bigr)$
denotes coordinates of the 
electrons belonging to the $I$-th group. Note that these electrons are not
confined to any particular region in the crystal.
The supercell hamiltonian $\hat H_S$ consists of single-particle terms
$\hat h$, which encompass kinetic energy as well as ionic and all
external potentials, and of an electron-electron contribution
\begin{multline}
\label{eq:vee_images}
\hat V_{ee}(\mathcal{R})=
 \sum_{j<i}\frac1{|\bi{r}_i-\bi{r}_j|}\\
 +\sum_i\Biggl[\frac12\sum_{j,\bi{R}_S\not=0}
   \frac1{|\bi{r}_i-\bi{r}_j-\bi{R}_S|}\Biggr]\,.
\end{multline}
The first term in \eref{eq:vee_images} represents the explicit Coulomb
interaction among electrons in the $N$-member group, and the second
term mimics the interactions with the electrons outside the group.
The physical meaning of the latter term is
as follows: A set of images is associated with each electron $j$,
and these virtual particles are placed at positions
$\bi{r}_j+\bi{R}_S$ so that they create a regular lattice. The combination
of all images has the same average charge density as the original
crystal and thus represents a reasonable environment for the selected
$N$ electrons. Each electron $i$ then interacts with the arrays of
charges associated with the other electrons in the group as well as with
its own images. Only one half of the interaction energy with images is
included in \eref{eq:vee_images}, the other half belongs to the rest of the
system and is distributed among the other members of the sum in
\eref{eq:supercell_ham_sum}. The model hamiltonian $\hat H_{\rm
  model}$ approaches the original fully interacting hamiltonian as $N$
increases and a larger fraction of the interactions has the exact form.

\begin{SCfigure*}
\resizebox{1.2\columnwidth}{!}{%
\includegraphics{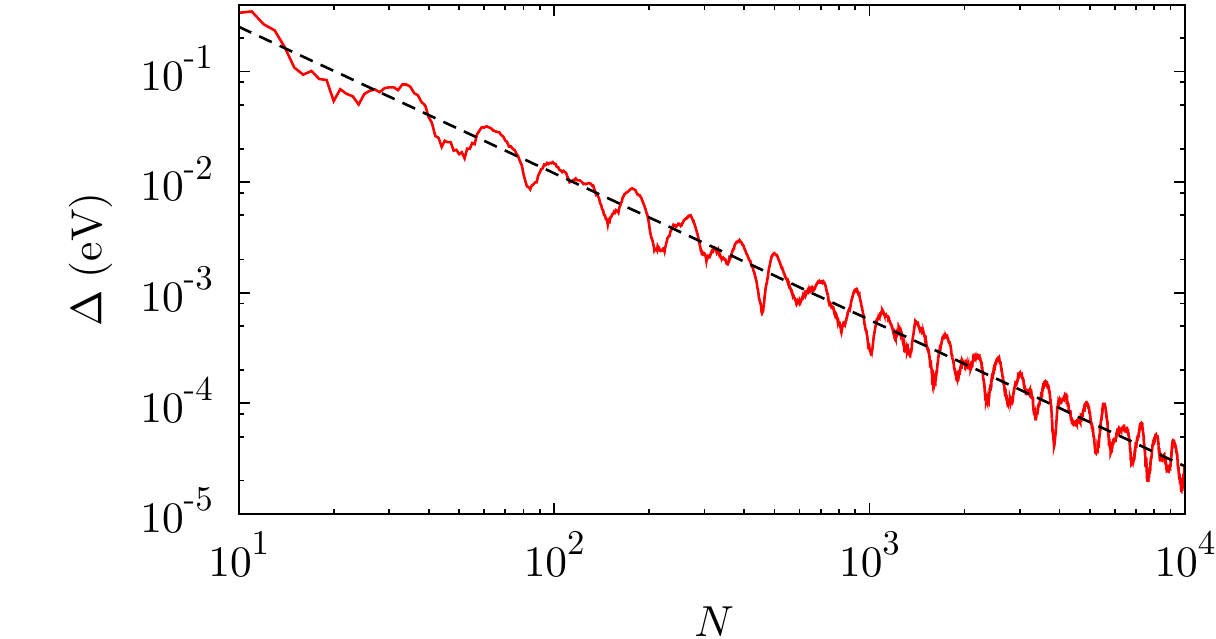}}
\caption{\label{fig:twist_av_free_gas}Deviation of the twist-averaged
  total energy \eref{eq:twist_av_energy} from the exact thermodynamic
  limit $E_{\infty}$, $\Delta=[E_S(N)-E_{\infty}]/N$, for a
  three-dimensional gas of non-interacting electrons with dispersion
  relation $\epsilon_{\bi{k}}=k^2/2$ at density $\rho$ corresponding to
  $r_{\rm s}=[3/(4\pi\rho)]^{1/3}=1$. The dashed line represents the
  average asymptotic decay of $\Delta$ that behaves as $N^{-1.32}$.}
\end{SCfigure*}

Hamiltonians $\hat H_S$ and $\hat H_{\rm model}$ possess the
symmetry described by \eref{eq:ham_transl_lda} with lattice
translations $\bi{R}$ replaced with $\bi{R}_S$. Consequently, the
eigenfunctions of $\hat H_S$ are many-particle Bloch waves 
\begin{equation}
\label{eq:MB_Bloch}
\Psi_{\bi{K}\alpha}(\mathcal{R})
 =U_{\bi{K}\alpha}(\mathcal{R})
  \exp\biggl(\rmi\bi{K}\cdot\sum_{i=1}^{N}\bi{r}_i\biggr)\,,
\end{equation}
where $\alpha$ is a many-body analogue of the band index and $\bi{K}$
is the crystal momentum \cite{rajagopal1994,rajagopal1995}. The wave
functions of the form \eref{eq:MB_Bloch} can be found in the same way
as the single-particle Bloch waves within the independent-particle
methods---as solutions to a problem of $N$ particles confined to a
simulation cell defined by vectors $\bi{L}_1$,
$\bi{L}_2$ and $\bi{L}_3$ belonging to the set $\{\bi{R}_S\}$ and
giving
$\Omega_S=\bigl|\bi{L}_1\cdot(\bi{L}_2\times\bi{L}_3)\bigr|$. The
dynamics of 
this finite $N$-particle system is governed by the hamiltonian $\hat H_S$
complemented with the so-called twisted boundary conditions \cite{lin2001}
\begin{multline}
\label{eq:twisted_BC}
\Psi_{\bi{K}\alpha}(\bi{r}_1,\dots,\bi{r}_i+\bi{L}_m,\dots,\bi{r}_N)
\\ =
 \Psi_{\bi{K}\alpha}(\bi{r}_1,\dots,\bi{r}_i,\dots,\bi{r}_N)
 \,\rme^{\rmi\bi{K}\cdot\bi{L}_m}\,,
\end{multline}
where $i=1,\dots,N$ and $m=1,2,3$. The indistinguishability of
electrons implies that the phase factor is the same irrespective of
which electron is moved, and therefore only a single $K$ vector common
to all electron coordinates is allowed in \eref{eq:MB_Bloch} and
\eref{eq:twisted_BC}. Once the quantum-mechanical problem in the
finite simulation cell is solved, wave functions for the whole
crystal can be constructed. Since there are no interactions between
individual $N$-particle groups, these wave functions have the form of an
antisymmetrized product of the Bloch functions \eref{eq:MB_Bloch}, namely
\begin{equation}
\label{eq:wf_whole_crystal}
\Psi_{\{\bi{K}_I\}\{\alpha_I\}}\bigl(\{\bi{r}\}\bigr)=
  \mathcal{A}\biggl[\prod_{I=1}^{\infty}\Psi_{\bi{K}_I\alpha_I}
  \bigl(\mathcal{R}^{(I)}\bigr)\biggr]\,.
\end{equation}
The indicated antisymmetrization is straightforward as long as all $\bi{K}_I$
in the product are different, because each factor $\Psi_{\bi{K}_I\alpha_I}$
then comes from a disjoint part of the Fock space. The total
energy corresponding to the wave function \eref{eq:wf_whole_crystal}
with the extra restriction $\bi{K}_I\not=\bi{K}_{I'}$ reads
\begin{equation}
\label{eq:energy_whole_crystal}
E_{\{\bi{K}_I\not=\bi{K}_{I'}\}\{\alpha_I\}}=\sum_{I=1}^{\infty}
\langle\Psi_{\bi{K}_I\alpha_I} |\hat H_S|\Psi_{\bi{K}_I\alpha_I}\rangle
\end{equation}
and the lowest energy is obtained by setting $\alpha_I=0$, that is, by
selecting the ground state for each of the different boundary
conditions. Although unlikely, it is possible that the true ground
state of $\hat H_{\rm model}$ falls outside the constraint
$\bi{K}_I\not=\bi{K}_{I'}$. In those cases, the expression
\eref{eq:energy_whole_crystal} with $\alpha_I=0$ is
an upper-bound estimate of the actual ground-state energy of $\hat H_{\rm
  model}$ with a bias diminishing as
$N$ increases. Taking into account the continuous character of the 
momentum $\bi{K}$ in the infinite crystal and the fact that all
possible boundary conditions \eref{eq:twisted_BC} are exhausted by
all $K$ vectors within the first Brillouin zone, the ground-state
energy per simulation cell can be written as
\begin{equation}
\label{eq:twist_av_energy}
E_S=\langle\hat H_S\rangle\equiv
 \frac{\Omega_S}{(2\pi)^3}\!\int\limits_{\rm 1. B.Z.}\!\rmd^3 K\,
 \langle\Psi_{\bi{K}0} |\hat H_S|\Psi_{\bi{K}0}\rangle\,.
\end{equation}
The total energy as well as expectation values of other periodic
operators are calculated as an average over all twisted boundary
conditions \cite{gammel1993,lin2001}. In practice, the
integral in \eref{eq:twist_av_energy} is approximated by
a discrete sum. The larger the simulation cell the smaller number of
$K$ points is needed to represent the integral, since the first
Brillouin zone of the simulation cell shrinks and the
$K$-dependence of the integrand gets weaker with increasing
$\Omega_S$.

Formula \eref{eq:twist_av_energy} is {\itshape almost} identical to
the expression used in supercell calculations
within the independent-particle theories, the only difference is
that the number of electrons at each $K$ point is fixed to
$N$. This constraint is benign in the case of insulators where
the number of occupied bands is indeed constant across the Brillouin
zone. In metals, however, the number of occupied bands fluctuates
from one $K$ point to another, and therefore the average
\eref{eq:twist_av_energy} does not coincide with the exact thermodynamic
limit. Figure~\ref{fig:twist_av_free_gas} provides an illustration of
the residual error. In
principle, it is possible to remove this error with the aid of
the grand-canonical description of the simulation cell \cite{lin2001},
but this concept is not straightforward to apply since the supercell is
no longer charge neutral.

% The supercell is
% no longer charge neutral and the final energy data contain extra
% statistical noise due to particle-number fluctuations that can
% potentially outweigh the benefits of the grand-canonical method
% altogether~\cite{drummond2008}.

\subsection{Ewald formula}
\label{sec:ewald}

Our definition of the simulation-cell hamiltonian~$\hat H_S$ in the
preceding section was only
formal and deserves a further commentary. It turns out that
$\hat V_{ee}$ is not absolutely convergent, and therefore it does not
unambiguously specify the interaction energy. In particular, the
seemingly periodic form of the sums in \eref{eq:vee_images} does not
by itself imply the desired periodicity of the hamiltonian. However,
enforcing this periodicity as an additional constraint makes the
definition unique and the resulting quantity is known as the Ewald
energy $\hat V_{ee}^{\rm (E)}$. It can be shown that the requirement of
periodicity is equivalent to a particular boundary condition imposed
on the electrostatic potential at infinity
\cite{deleeuw1980a,fraser1996}. The peculiar convergence properties of
the sums in \eref{eq:vee_images} are a manifestation of the long-range
character of the Coulomb potential---the boundary of the sample is
never irrelevant, no matter how large its volume is. Consequently,
careful considerations are required in order to perform the
thermodynamic limit correctly.

For the purposes of practical evaluation in QMC simulations, the Ewald
energy is written as
\begin{multline}
\label{eq:ewald_energy}
\hat V_{ee}^{\rm (E)}(\mathcal{R})=
  \sum_{j<i} V_{\rm E}(\bi{r}_i-\bi{r}_j)\\[-.3em]
  +\frac{N}2\lim_{r\to 0}\biggl(
   V_{\rm E}(\bi{r})-\frac1{|\bi{r}|}\biggr)\,,
\end{multline}
where $V_{\rm E}(\bi{r}_i-\bi{r}_j)$ stands for an electrostatic
potential at $\bi{r}_i$
induced by the charge at $\bi{r}_j$ together with its images located
at $\bi{r}_j+\bi{R}_S$. An explicit formula for
the Ewald pair potential $V_{\rm E}$ reads \cite{allen1989,deleeuw1980a}
\begin{align}
\label{eq:ewald_potential}
&V_{\rm E}(\bi{r}_i-\bi{r}_j)
\nonumber\\
&\hskip 2em
=\sum_{\bi{R}_S}\frac1{|\bi{r}_i-\bi{r}_j-\bi{R}_S|}
    \erfc\bigl(\kappa|\bi{r}_i-\bi{r}_j-\bi{R}_S|\bigr)
\nonumber\\
&\hskip 2em
\phantom{=} +\frac{4\pi}{\Omega_S}\sum_{\bi{G}_S\not =0}\frac1{G_S^2}
    \exp\biggl[-\frac{G_S^2}{4\kappa^2}
   +\rmi\bi{G}_S\cdot(\bi{r}_i-\bi{r}_j)\biggr]
\nonumber\\
&\hskip 2em
\phantom{=} -\frac{\pi}{\Omega_S\kappa^2}\,,
\end{align}
where $\bi{G}_S$ are vectors reciprocal to $\bi{R}_S$, $\exp(\rmi
\bi{G}_S\cdot\bi{R}_S)=1$, and $\kappa$ is an arbitrary positive
constant that does not alter the value of $V_{\rm E}$.
% The ``self-interaction'' $\xi$ can be further expressed as
% \begin{eqnarray}
% \label{eq:madelung_potential}
% \fl
% \xi=\lim_{\bi{r}_i\to\bi{r}_j}\biggl(
%    V_{\rm E}(\bi{r}_i-\bi{r}_j)-\frac1{|\bi{r}_i-\bi{r}_j|}\biggr)
% =\sum_{\bi{R}_S\not =0}\frac{\erfc(\alpha R_S)}{R_S}
%   -\frac{2\alpha}{\sqrt{\pi}}\nonumber\\
%  +\frac{4\pi}{\Omega_S}\sum_{\bi{G}_S\not =0}\frac1{G_S^2}
%     \exp\biggl(-\frac{G_S^2}{4\alpha^2}\biggr)
%  -\frac{\pi}{\Omega_S\alpha^2}\,.
% \end{eqnarray}
The omission of the $\bi{G}_S=0$ term in the reciprocal sum
corresponds to the removal of the homogeneous component from the
potential $V_{\rm E}$. When evaluating the total energy of a charge neutral
crystal, these ``background'' contributions exactly cancel among the Ewald
energies of electron-electron, electron-ion and ion-ion interactions.

The important feature of the Ewald formula \eref{eq:ewald_potential}
is the decomposition of the slowly converging Coulomb sum into two
rapidly converging parts, one in real space
and the other in reciprocal space. The break-up is not unique (not
only due to the arbitrariness of~$\kappa$) and can be further
optimized for computational efficiency \cite{natoli1995a,rajagopal1994b}.

\subsection{Extrapolation to the thermodynamic limit}
\label{sec:fse_extrapolation}

The total energy per electron $\epsilon_N=E_S/N$ evaluated according
to the outlined recipe still depends on the size of the simulation
cell. These residual finite-size effects can be removed by
an extrapolation: Energy~$\epsilon_N$ is calculated in simulation
cells of several sizes and an appropriate function
$\epsilon^{\rm fit}(N)=\epsilon_{\infty}+f(N)$ is subsequently fitted
through the acquired data. In the end, $\epsilon_{\infty}$ is the desired
energy per electron in the thermodynamic limit, where the error term $f(N)$
by definition vanishes, $\lim_{N\to\infty}f(N)=0$. Experience with a
wide range of 
systems \cite{ceperley1978,ceperley1987,pozzo2008b,kolorenc2008}
indicates that as long as the integral over the Brillouin zone in
\eref{eq:twist_av_energy} is well converged, the finite-size data are
well approximated by a smooth function $f(N)<0$ dominated by a $1/N$
contribution.%
\footnote{The finite-size scaling depends on the dimensionality of the
  problem and the $1/N$ dependence corresponds to three-dimensional
  samples considered here.}
The size extrapolation is therefore quite straightforward,
although often computationally expensive due to the relatively slow
decay of the error term.

The origin and behaviour of the finite-size effects is a subject of
ongoing investigations with the aim to find means of accelerating the
convergence of the total energy and other expectation values to the
thermodynamic limit. Furthermore, understanding the dependence of
the finite-size errors on various 
parameters, such as particle density, can reduce the number of
explicit size extrapolations needed to obtain quantities of
interest. In calculations of equations of state
(section~\ref{sec:app_eos}), for instance, it 
is then sufficient to perform the extrapolation only at
selected few, instead of all, electron densities \cite{kolorenc2008}.

It turns out that, in the twist-averaged expectation values
$\langle\hat H_S\rangle$ calculated
in finite simulation cells, both the hamiltonian and the wave function
contain biases that contribute to the 
$1/N$ asymptotics of the error term $f(N)$. It was argued
\cite{fraser1996,williamson1997} that the slow converging
parts of the hamiltonian reside in the
exchange-correlation energy
\begin{equation}
\label{eq:xc_energy}
V_{\rm XC}=\langle \hat V_{ee}^{\rm (E)}\rangle
 - \frac12 \kern-.9em\int\limits_{\Omega_S\times\Omega_S}\kern-.9em \rho(\bi{r})
  V_{\rm E}(\bi{r}-\bi{r}')\rho(\bi{r}')\,\rmd^3 r\rmd^3 r'
\end{equation}
defined as a difference between the expectation value of the
Ewald energy $\langle \hat V_{ee}^{\rm (E)}\rangle$ and the Hartree term that
describes the interaction of the charge densities
$\rho(\bi{r})=\langle\hat\rho(\bi{r})\rangle$. The Hartree energy
is found to converge rather rapidly with the size of the
simulation cell.
In systems with cubic and higher symmetry, the leading contribution to 
$f(N)$ can be written as \cite{chiesa2006}
\begin{align}
\label{eq:fse_xc}
f_{\rm XC}(N)
  &=\frac{V_{\rm XC}}{N}-\lim_{N\to\infty}\frac{V_{\rm XC}}{N}
\nonumber\\[.2em]
  &= -\frac{2\pi}{\Omega_S}\lim_{G_S\to 0}\frac{S(\bi{G}_S)}{G_S^2}\,.
\end{align}
This formula involves the static structure factor
\begin{equation}
\label{eq:sk}
S(\bi{G}_S)=\frac1N\Bigl[\langle\hat\rho(\bi{G}_S)\hat\rho(-\bi{G}_S)\rangle
-\bigl|\langle\hat\rho(\bi{G}_S)\rangle\bigr|^2
\Bigr]\,,
\end{equation}
where $\hat\rho(\bi{G}_S)$ is a Fourier component of the density
operator.
%Note that since the charge density $\rho$ has a periodicity of the
%primitive cell, the second term in \eref{eq:sk} contributes only
%for those $G_S$ that coincide with reciprocal vectors corresponding to
%the primitive lattice.
The exact small-momentum asymptotics of the
structure factor in Coulomb systems is given by the random phase
approximation (RPA) \cite{bohm1953} and reads $S(\bi{G}_S)\sim G_S^2$,
which ensures that the limit in~\eref{eq:fse_xc} is well
defined. In systems with lowered symmetry and for less
accurate approximations, as is the Hartree--Fock theory where
$S(\bi{G}_S)\sim G_S$, the
expression for $f_{\rm XC}(N)$ must be appropriately modified
\cite{drummond2008}.

The random phase approximation provides insight also into the
finite-size effects induced by restricting the wave function into a
finite simulation cell. According to the RPA, the many-body wave
function in Coulomb systems factorizes as
\begin{equation}
\label{eq:wf_RPA_asympt}
\Psi(\mathcal{R})=\Psi_{\rm s.r.}(\mathcal{R}) \exp
  \biggl[-\sum_{j<i} u(\bi{r}_i-\bi{r}_j)\biggr]\,,
\end{equation}
where $\Psi_{\rm s.r.}$ contains only short-range correlations and
the function $u(\bi{r})$ decays as $1/r$ at large
distances. Such
 long-range behaviour is not consistent with the boundary
conditions \eref{eq:twisted_BC}, and a truncation of this tail is
therefore unavoidable. The corresponding finite-size bias is most
pronounced in the expectation value of the kinetic energy
$T=\langle\hat T\rangle$ and contributes an error term \cite{chiesa2006}
\begin{align}
\label{eq:fse_kinetic}
f_T(N)&=\frac{T}{N}-\lim_{N\to\infty}\frac{T}{N}
\nonumber\\[.2em]
 &=-\frac{1}{4\Omega_S}\lim_{G_S\to 0} G_S^2\, u(\bi{G}_S)\,,
\end{align}
where $u(\bi{G}_S)\sim 1/G_S^2$ is a Fourier component of $u(\bi{r})$.

Assuming that we are able to evaluate the expressions
\eref{eq:fse_xc} and \eref{eq:fse_kinetic}, 
we can decompose $f(N)$ into parts
$f(N)=f_{\rm XC}(N)+f_T(N)+f'(N)$, 
where $f'(N)$ is substantially smaller than $f(N)$ and the
size extrapolation is therefore better controlled. In the case of the
homogeneous electron gas, the RPA provides analytic expressions for
the small momentum behaviour of the required quantities
$S(\bi{G}_S)\approx G_S^2/(2\omega_{\rm p})$ and
$u(\bi{G}_S)\approx 4\pi/(G_S^2\omega_{\rm p})$, where
$\omega_{\rm p}=\sqrt{4\pi N/\Omega_S}$ is the plasma frequency. Subsequently,
the individual error terms simplify to
\begin{equation}
\label{eq:fse_heg}
f_{\rm XC}(N)=-\frac1N\, \frac{\omega_{\rm p}}{4}
\quad\hbox{and}\quad
f_{\rm T}(N)=-\frac1N\, \frac{\omega_{\rm p}}{4}\,.
\end{equation}
It can be demonstrated that these two
contributions completely recover the $1/N$ part of $f(N)$, and that the
residual term $f'(N)$ scales as $\sim N^{-4/3}$ \cite{drummond2008}.

Application of the derived finite-size corrections to simulations of
realistic solids is less straightforward
 since reliable analytic
results are not available. The small momentum asymptotics of
$S(\bi{G}_S)$ and $u(\bi{G}_S)$ have to be examined 
numerically. Sufficiently large simulation cells are needed for this
purpose, since the smallest nonzero reciprocal vector available in a
supercell with volume $\Omega_S$ is $G_S\sim\Omega_S^{-1/3}$.
Utilization of the kinetic energy correction
\eref{eq:fse_kinetic} within DMC simulations is further complicated by
the fact that the function $u(\bi{r})$ is not given as an expectation value of
an operator, and thus it is not clear how it could be extracted from
the sampled mixed distribution $\Psi_{\rm D}^*\Psi_{\rm T}$. One has to
rely on the trial wave function alone to correctly reproduce the long-range
tail \eref{eq:wf_RPA_asympt}, which can be a challenging task in
simulation cells containing a large number of electrons.

The above analysis employs exact analytic formulae or quantum Monte Carlo
simulations themselves to find corrections to the finite-size biases.
Alternatively, it is possible to adopt a more heuristic
approach and estimate the finite-size effects within an
approximative method. For instance, the error term $f(N)$ can be
rewritten as
$f(N)=\epsilon^{\rm (LDA)}_S-\epsilon^{\rm (LDA)}_{\infty}+f''(N)$,
where $\epsilon^{\rm (LDA)}_S$ and $\epsilon^{\rm (LDA)}_{\infty}$ are
total energies per particle provided by the local density
approximation with two distinct exchange-correlation functionals,
and $f''(N)$ is anticipated to be considerably smaller than
$f(N)$~\cite{kwee2008}. The exchange-correlation functional
corresponding to 
$\epsilon^{\rm (LDA)}_{\infty}$ is constructed from properties of the
homogeneous electron gas in the thermodynamic limit (in other words, it is the
standard LDA functional), the functional leading to $\epsilon^{\rm
  (LDA)}_S$ is based on the homogeneous electron gas confined to the
same finite supercell as the quantum system under 
investigation. The latter functional is not universal and needs
to be found for each simulation cell separately at the cost
of auxiliary simulations of the homogeneous Coulomb gas.

\subsection{An alternative model for Coulomb interaction energy}

The expression for the electron-electron interaction energy $\hat
V_{ee}^{\rm (E)}$ has two properties: (i) it is periodic and (ii)
corresponds to an actual, albeit artificial, system of point
charges. Although the latter property is conceptually convenient, it
is not really necessary, and any periodic potential that
exhibits the correct behaviour in the limit of the infinitely large
simulation cell is legitimate. Relaxation of the constraint (ii) in
favour of faster convergence of the total energy per
particle~$\epsilon_N$ to its thermodynamic limit was explored in a
series of papers 
\cite{fraser1996,williamson1997,kent1999}, where the so-called model
periodic Coulomb (MPC) interaction was proposed. The replacement for
the Ewald energy $\hat V_{ee}^{\rm (E)}$ reads
\begin{align}
\label{eq:mpc_energy}
&\hat V_{ee}^{(\rm MPC)}(\mathcal{R})
=\sum_{j<i} \frac1{|\bi{r}_i-\bi{r}_j|_{\rm m}}\\[.2em]
&\hskip1.5em
  +\sum_i\int\limits_{\Omega_S}\biggl[
    V_{\rm E}(\bi{r}_i-\bi{r})-\frac1{|\bi{r}_i-\bi{r}|_{\rm m}}\biggr]
    \rho(\bi{r})\,\rmd^3r\nonumber\\[.2em]
&\hskip1.5em
  -\kern-.9em \int\limits_{\Omega_S\times\Omega_S}\kern-.5em\biggl[
    V_{\rm E}(\bi{r}-\bi{r}')-\frac1{|\bi{r}-\bi{r}'|_{\rm m}}\biggr]
    \rho(\bi{r})\rho(\bi{r}')\,\rmd^3r\rmd^3r'\,,
\nonumber
\end{align}
where $|\bi{r}-\bi{r}'|_{\rm
  m}=\min_{\bi{R}_S}|\bi{r}-\bi{r}'+\bi{R}_S|$ stands for the
so-called minimum image distance. The operator~$\hat V_{ee}^{(\rm
  MPC)}$ is constructed in
such a way that the Hartree part of its expectation value is the same as in
the Ewald method, whereas the slowly converging contribution to the
exchange-correlation energy is removed. Therefore,
the MPC interaction is essentially equivalent to the Ewald formula
\eref{eq:ewald_energy} complemented with the {\em a posteriori\/}
correction
\eref{eq:fse_xc}. Instead of the structure factor, it is the
one-particle density $\rho$ that has to be evaluated as an extra
quantity in this case (unless it is known exactly as in a homogeneous
system). The explicit presence of the density $\rho$ in the
hamiltonian is inconvenient for the DMC method where the local energy
is needed from the start of the simulation, that is, before the density
data could be accumulated. The situation can be remedied by replacing the
unknown density $\rho$ with an approximation $\rho_A$. For
instance, the one-particle density provided by DFT is usually quite
accurate. The error introduced by this substitution is proportional to
$(\rho-\rho_A)^2$ and further diminishes as the
simulation cell increases. The Ewald and MPC energies per
particle are therefore the same in the thermodynamic limit even if the
approximate charge density is used
\cite{maezono2003,drummond2008}.
% Alternatively, the DMC Green's 
% function can be evaluated with the Ewald energy. In fact, the
% equivalence of the MPC interaction with the corrected Ewald method
% corresponds to just this case when the DMC projection is performed
% with the Ewald hamiltonian. 

\section{Trial wave functions}
\label{sec:wf}

Accurate trial wave functions are essential for successful
applications of the quantum Monte Carlo methods. The quality of the employed
wave functions influences 
the statistical efficiency of the simulations as well as the accuracy
of the achieved results. Equally important, especially for
investigations of extended systems, is the possibility to quickly
compute the wave function value and its
derivatives ($\nabla\Psi_{\rm T}$ and $\nabla^2\Psi_{\rm T}$), since
these quantities usually represent the most computationally costly
part of the whole simulation. Compact expressions are therefore
strongly preferred.

A significant part of the construction of the trial wave functions is
optimization of the variational parameters introduced into the
functional form representing~$\Psi_{\rm T}$. It is a non-trivial task
since the number of parameters is often large, $\Psi_{\rm T}$ depends
non-linearly on them, and the quantity to be
minimized ($E_{\rm VMC}$ or the variance of the local energy) is a
fluctuating number. Several powerful methods addressing these
problems have been developed during the years
\cite{nightingale2001,umrigar2005,toulouse2007a,umrigar2007} and even
hundreds of parameters can be optimized with good efficiency nowadays.

\subsection{Elementary properties}
\label{sec:wf_elementary}

Since our aim is the electronic structure, and the electrons are
subject to the Pauli exclusion principle, our trial wave functions have
to be antisymmetric with respect to pair-electron exchanges. We
assume collinear spins that are independent of electron positions, and
therefore the full wave function $\widetilde\Psi_{\rm T}$ can be
factorized as
\begin{multline}
\label{eq:wf_spin_factorization}
\widetilde\Psi_{\rm T}(\mathcal{R},\mathcal{S})
=\sqrt{\frac{N_{\uparrow}!\,N_{\downarrow}!}{N!}}\,
 \sum_C (-1)^C \\ %\mathop{\rm sign}(C)
 \times\Psi_{\rm T}(C\mathcal{R})\,
 \bigl|C\{\underbrace{\uparrow\ldots\uparrow}_{\displaystyle N_{\uparrow}}
     \underbrace{\downarrow\ldots\downarrow}_{\displaystyle N_{\downarrow}}\}
 \bigr\rangle\,,
\end{multline} 
where $\mathcal{S}=(\sigma_1,\ldots,\sigma_N)$ is a vector consisting of
$N=N_{\uparrow}+N_{\downarrow}$ spin variables. The sum runs over all
distinct states of $N_{\uparrow}$
up-oriented and $N_{\downarrow}$ down-oriented spins. In
the case of $N_{\uparrow}=2$ and $N_{\downarrow}=1$ the spin states
are $|\!\uparrow_1\uparrow_2\downarrow_3\rangle$,
$|\!\uparrow_1\uparrow_3\downarrow_2\rangle$ and
$|\!\uparrow_2\uparrow_3\downarrow_1\rangle$, and the corresponding
$C\mathcal{R}$ combinations are $\{\bi{r}_1,\bi{r}_2,\bi{r}_3\}$,
$\{\bi{r}_1,\bi{r}_3,\bi{r}_2\}$ and $\{\bi{r}_2,\bi{r}_3,\bi{r}_1\}$.
The spatial-only part $\Psi_{\rm T}$ is
antisymmetric with respect to exchanges of parallel electrons and
its symmetry with respect to exchanges of antiparallel electrons is
unrestricted. The sum in~\eref{eq:wf_spin_factorization} with the
appropriate sign factors $(-1)^C$ represents the residual
antisymmetrization for the antiparallel spins.

Both $\widetilde\Psi_{\rm T}$ and $\Psi_{\rm T}$ are normalized 
to unity and identity
$\langle \widetilde\Psi_{\rm T}|\hat A|\widetilde\Psi_{\rm T}\rangle=
\langle \Psi_{\rm T}|\hat A|\Psi_{\rm T}\rangle$ holds for any
spin-independent operator $\hat A$ 
 since the spin states
$\bigl|C\{\uparrow\ldots\uparrow\downarrow\ldots\downarrow\}\bigr\rangle$
are mutually orthonormal. Therefore, it is usually sufficient to
consider only the spatial part~$\Psi_{\rm T}$ of the full many-body
wave function in applications of the VMC
and DMC methods, and we limit our discussion to $\Psi_{\rm T}$ from now on.%
\footnote{In fact, the DMC algorithm is defined only for the spatial
  part $\Psi_{\rm T}$, consult sections~\ref{sec:dmc_sampling}
  and~\ref{sec:dmc_spins}. Decomposition
  \eref{eq:wf_spin_factorization} then provides a hint how to
  calculate expectation values of spin-dependent operators from the
  sampled mixed distribution $\Psi_{\rm D}^*\Psi_{\rm T}$.}

Our goal is for the local energy $\hat H\Psi_{\rm T}/\Psi_{\rm T}$ to be
very close to a hamiltonian eigenvalue and fluctuating as little as
possible. In systems with charged particles interacting via the Coulomb
potentials, it requires that the kinetic energy proportional to
$\nabla^2\Psi_{\rm T}$ contains singularities 
which cancel the $1/r$
divergencies of the potential. This cancellation is vital for
controlling the statistical uncertainties of the Monte Carlo estimate
of the total energy and takes place when Kato cusp
conditions are fulfilled \cite{kato1957,pack1966}.

At electron-electron coincidences, these conditions can be formulated
with the aid of projections of the trial wave function
$\Psi_{\rm T}$ onto spherical harmonics $Y_{lm}$ centered at the
coincidence point,
\begin{multline}
\Psi_{\rm T}^{(l,m)}
(r_{ij},\bi{r}_{\rm c.m.},\mathcal{R}\setminus\{\bi{r}_i,\bi{r}_j\})\\
=\frac1{r_{ij}^l}\int_{4\pi}
 \Psi_{\rm T}(\mathcal{R})
Y^*_{lm}(\Omega_{ij})\rmd\Omega_{ij}\,.
\end{multline}
% \begin{equation}
% \Psi_{\rm T}(\mathcal{R})=\sum_{l=0}^{\infty}\sum_{m=-l}^l
% r_{ij}^l \Psi_{\rm T}^{(l,m)} Y_{lm}\biggl(\frac{\bi{r}_{ij}}{r_{ij}}\biggr)
% \end{equation}
In this definition, the following notation has
been introduced: $r_{ij}=|\bi{r}_{ij}|=|\bi{r}_i-\bi{r}_j|$ is the
electron-electron distance, $\Omega_{ij}$ is the spherical angle
characterizing orientation of the vector $\bi{r}_{ij}$, and $\bi{r}_{\rm
  c.m.}=(\bi{r}_i+\bi{r}_j)/2$ is the position of the center of mass
of the electron pair. The cusp conditions can then be written as
\begin{equation}
\label{eq:cusp_ud}
\lim_{r_{ij}\to 0}\frac1{\Psi_{\rm T}^{(0,0)}}\,
  \frac{\partial\Psi_{\rm T}^{(0,0)}}{\partial r_{ij}}=\frac12
\end{equation}
for unlike spins and
\begin{equation}
\label{eq:cusp_uu}
\lim_{r_{ij}\to 0}\frac1{\Psi_{\rm T}^{(1,m)}}\,
  \frac{\partial\Psi_{\rm T}^{(1,m)}}{\partial r_{ij}}=\frac14
\end{equation}
for like spins.
The expressions~\eref{eq:cusp_ud} and~\eref{eq:cusp_uu} differ because
$\Psi_{\rm T}$ is an odd function with respect to $\bi{r}_{ij}$ in the
latter case, which implies $\Psi_{\rm T}^{(0,0)}=0$.

Analogous cusps occur in all-electron calculations when electrons
approach nuclei. Unless the $s$-wave component $\Psi_{\rm T}^{(0,0)}$
vanishes (for a general discussion see \cite{pack1966}), it can be
shown that
\begin{equation}
\label{eq:cusp_nucl}
\lim_{r_{Ii}\to 0}\frac1{\Psi_{\rm T}^{(0,0)}}\,
  \frac{\partial\Psi_{\rm T}^{(0,0)}}{\partial r_{iI}}=-Z_I\,,
\end{equation}
where $r_{iI}$ is the electron-nucleus distance and $Z_I$ is the
charge of the nucleus.

A convenient functional form that meets the specified
criteria is a product of an antisymmetric part~$\Psi_{\rm A}$ and a
positive symmetric expression $\exp(-U_{\rm corr})$ \cite{jastrow1955},
\begin{equation}
\label{eq:wf_factorization}
\Psi_{\rm T}(\mathcal{R})=\Psi_{\rm A}(\mathcal{R})\,
\exp\bigl[-U_{\rm corr}(\mathcal{R})\bigr]\,.
\end{equation}
The Jastrow correlation factor $\exp(-U_{\rm corr})$ incorporates the
electron-electron cusp conditions~\eref{eq:cusp_ud}
and~\eref{eq:cusp_uu}, and $\Psi_{\rm A}$ ensures the fermionic
character 
of the wave function. The electron-ion cusps \eref{eq:cusp_nucl} can
be included in either $\Psi_{\rm A}$
\cite{manten2001,ma2005a,ma2005b,esler2010} or in the correlation
factor \cite{drummond2004}. In simulations of extended systems, the 
antisymmetric part obeys the twisted boundary
conditions (section~\ref{sec:tabc}) and $\exp(-U_{\rm corr})$ is
periodic at the boundaries of the simulation cell. We discuss the
individual parts of the trial wave function~\eref{eq:wf_factorization}
in some detail next.

\subsection{Jastrow factor}
\label{sec:wf_jastrow}

The majority of applications fits into a framework set by the
expression
\begin{equation}
\label{eq:wf_jastrow}
U_{\rm corr}(\mathcal{R})=\sum_i\chi_{\sigma_i}(\bi{r}_i) +
\sum_{j<i} u_{\sigma_i\sigma_j}(\bi{r}_i,\bi{r}_j)\,,
\end{equation}
where the functions $\chi$ and $u$ take a specific parametrized form
\cite{reynolds1982,umrigar1988,schmidt1990} and can depend
also on spins of the involved electrons as 
indicated by the indices $\sigma\in\{\uparrow,\downarrow\}$. The
inclusion of the uncorrelated one-body terms $\chi$ is important
especially if the trial wave function is optimized with a fixed
antisymmetric part $\Psi_{\rm A}$
\cite{fahy1990,foulkes2001,umezawa2003}. The two-body terms $u$ are typically
simplified to 
\begin{multline}
\label{eq:wf_jastrow_2body}
\sum_{j<i} u_{\sigma_i\sigma_j}(\bi{r}_i,\bi{r}_j) \to
\sum_{j<i} u_{\rm ee}(\bi{r}_{ij}) \\ + 
\sum_{j<i,I} u_{\rm een}
 (|\bi{r}_{ij}|,|\bi{r}_{iI}|,|\bi{r}_{jI}|)\,,
\end{multline}
where $u_{\rm ee}$ is an expression corresponding to a homogeneous
system and the electron-electron-nucleus term~$u_{\rm een}$ takes into
account the differences between the two-body correlations in
high-density regions near nuclei and in low density regions far from
them. Spin indices were dropped to simplify the notation. The $u_{\rm
een}$ contribution can usually be short ranged in the $|\bi{r}_{ij}|$
parameter, whereas the simpler $u_{\rm ee}$ term is preferably long
ranged in order to approximate the RPA asymptotics
\eref{eq:wf_RPA_asympt} as closely as allowed by the given simulation
cell \cite{ceperley1978,drummond2004}. Of course, limited
computational resources can (and often do) enforce further simplifying
compromises. In simulations of homogeneous fermion fluids (electron
gas, ${}^3$He), on the other hand, even higher order correlations were
successfully included: three-particle
\cite{lhuillier1981,lee1981,schmidt1981,kwon1993,kwon1998} as well as
four-particle \cite{holzmann2006}.

\subsection{Slater--Jastrow wave function}
\label{sec:wf_slater}

The simplest antisymmetric form that can be used in place of
$\Psi_{\rm A}$ in \eref{eq:wf_factorization} is a product of two
Slater determinants, 
\begin{align}
\label{eq:wf_slater}
\Psi_{\rm A}^{\rm Slater}(\mathcal{R})&=
\mathcal{A}\bigl[
\psi_1^{\uparrow}(\bi{r}_1^{\uparrow})\ldots
\psi_{N_{\uparrow}}^{\uparrow}(\bi{r}_{N_{\uparrow}}^{\uparrow})
\bigr]
\nonumber\\
&\hskip 3em
\times \mathcal{A}\bigl[
\psi_1^{\downarrow}(\bi{r}_1^{\downarrow})\ldots
\psi_{N_{\downarrow}}^{\downarrow}(\bi{r}_{N_{\downarrow}}^{\downarrow})
\bigr]\nonumber \\[.2em]
&=
\det\bigl[\psi_{n}^{\uparrow}(\bi{r}_i^{\uparrow})\bigr]
\det\bigl[\psi_{m}^{\downarrow}(\bi{r}_j^{\downarrow})\bigr]\,,
\end{align}
where $\psi_{n}^{\uparrow}$ and $\psi_{m}^{\downarrow}$ are
single-particle orbitals that correspond to spin-up respectively
spin-down electronic states and
$\psi_{n}^{\sigma}(\bi{r}_i^{\sigma})$ in the arguments of the
determinants stands for a $N_{\sigma}\times N_{\sigma}$
matrix $A_{n i}^{\sigma}$. In quantum-chemical applications, a
common strategy to improve upon the Slater wave function is to use a
linear combination of several determinants,
\begin{equation}
\label{eq:wf_multidet}
\Psi_{\rm A}^{\rm m\mhyphen det}(\mathcal{R})=\sum_{\alpha} c_{\alpha}
\det\bigl[\psi_{\alpha,n}^{\uparrow}(\bi{r}_i^{\uparrow})\bigr]
\det\bigl[\psi_{\alpha,m}^{\downarrow}(\bi{r}_j^{\downarrow})\bigr]\,.
\end{equation}
These multi-determinantal expansions are mostly
impractical for simulations of solids, since the number of
determinants required to describe the wave function to some fixed
accuracy increases exponentially with the system size. One case where
multiple determinants are vital even in extended systems are fixed-node DMC
calculations of excitation energies, since adherence to proper
symmetries is essential for validity of the corresponding variational
theorem \cite{foulkes1999,towler2000} and the trial wave functions
displaying the correct symmetry are not always representable by a
single determinant. In these instances, however, the expansions
\eref{eq:wf_multidet} are short.
%A possible exception are investigations of localized structures in
%crystals (such as point defects \cite{hood2003}) or on their surfaces.

In simulation cells subject to the twisted boundary conditions
\eref{eq:twisted_BC} specified by a supercell crystal momentum
$\bi{K}$, the one-particle orbitals $\psi^{\sigma}_m$ are Bloch waves
satisfying
\begin{equation}
\label{eq:bloch_supercell}
\psi_{\bi{K}m}^{\sigma}(\bi{r}+\bi{L}_{\alpha})
 =\psi_{\bi{K}m}^{\sigma}(\bi{r})\,
 \rme^{\rmi\bi{K}\cdot\bi{L}_{\alpha}}\,,
\end{equation}
where $\alpha=1,2,3$ and $m$ is a band index in the
supercell. Since the simulation cell contains several primitive cells,
the crystal has a higher translational symmetry than implied by
\eref{eq:bloch_supercell} and the orbitals can be conveniently
re-labeled using $m\to(\bi{k},m')$, where $\bi{k}$ and~$m'$ are a
momentum and a band index defined with respect to the
primitive cell. The Bloch waves fulfill also
\begin{equation}
\label{eq:bloch_primcell}
\psi_{\bi{K}\bi{k}m'}^{\sigma}(\bi{r}+\bi{l}_{\alpha})
 =\psi_{\bi{K}\bi{k}m'}^{\sigma}(\bi{r})\,
 \rme^{\rmi\bi{k}\cdot\bi{l}_{\alpha}}\,,
\end{equation}
where $\bi{l}_{\alpha}$ are lattice vectors defining the primitive
cell. Assuming that the supercell is built as
$\bi{L}_{\alpha}=n_{\alpha}\bi{l}_{\alpha}$ with integers $n_{\alpha}$,
the momenta $\bi{k}$ compatible with \eref{eq:bloch_supercell} fall
onto a regular mesh
\begin{multline}
\label{eq:pack_monkhorst}
\fl
\bi{k}=\bi{K}+2\pi\biggl(
 \frac{j_1}{n_1}\frac{\bi{l}_2\times\bi{l}_3}{%
   \bi{l}_1\cdot(\bi{l}_2\times\bi{l}_3)} \\
+\frac{j_2}{n_2}\frac{\bi{l}_3\times\bi{l}_1}{%
   \bi{l}_2\cdot(\bi{l}_3\times\bi{l}_1)}
+\frac{j_3}{n_3}\frac{\bi{l}_1\times\bi{l}_2}{%
   \bi{l}_3\cdot(\bi{l}_1\times\bi{l}_2)}
\biggr)
\end{multline}
with indices $j_{\alpha}$ running from 0 to $n_{\alpha}-1$. This set
of $k$ points corresponds to the Monkhorst--Pack grid \cite{monkhorst1976}
shifted by a vector $\bi{K}$.

A number of strategies have been devised
to determine the optimal one-particle orbitals for use in the
Slater--Jastrow wave functions, which certainly differ from the 
Hartree--Fock orbitals that minimize the variational energy only when
$U_{\rm corr}=0$. Ideally, the orbitals are parametrized by an 
expansion in a saturated basis with the expansion coefficients varied
to minimize the VMC or DMC total energy. The stochastic
noise and the computational demands of the DMC method make
the minimization of $E_{\rm DMC}$ extremely inefficient in practice. 
The VMC optimization of the orbitals was
successfully performed in atoms and small molecules of the first-row
atoms \cite{filippi1996,umrigar2007,toulouse2008}, but the method is
still too demanding for applications to solids.

To avoid the large number of variational parameters needed to describe
the single-particle orbitals, another family of methods has been
proposed. The orbitals in the Slater--Jastrow wave function are found
as solutions to self-consistent-field equations that represent a
generalization of the Hartree--Fock theory to the presence of the
Jastrow correlation factor
\cite{filippi2000,schautz2004,umezawa2003,umezawa2006}. These
methods were tested in atoms as well as in
solids within the VMC framework
\cite{filippi2000,sakuma2006}. Unfortunately, the wave functions 
derived in this way did not 
lead to lower DMC energies compared to wave functions with orbitals
from the Hartree--Fock theory or from the local density approximation
\cite{filippi2000,prasad2007}. It is unclear, whether the lack of
observed improvements in the fermionic nodal surfaces stems from
insufficient flexibility of the employed correlation factors or from
the fact that only applications to weakly correlated systems were
considered so far.

An even simpler approach is to introduce a parametric dependence into the
self-consistent-field equations without a direct relation to the
actual Jastrow factor. An example are the Kohn--Sham equations
corresponding to some exchange-correlation functional, in which it is
possible to identify a parameter (or several parameters) measuring the
degree of correlations in the system and thus mimicking, to a certain
degree, the effect of the Jastrow factor. Particular hybrid
functionals with variable admixture of the exact-exchange component
\cite{kummel2008} were successfully employed for this purpose in
conjunction with the DMC optimization, so that the variations of the
fermionic nodal structure could be directly quantified
\cite{wagner2003,wagner2007a,sola2009,kolorenc2010}. Sizeable
improvements of the DMC total energy associated with the
replacement of the Hartree--Fock (or LDA) orbitals with the orbitals
provided by the optimal hybrid functional were observed in compounds
containing $3d$ elements.

Evaluation of the Slater determinants dominates the
computational demands of large-scale Monte Carlo
calculations, and it is therefore very important to consider its
implementation carefully. Notably, schemes combining a localized basis set
(atom-centered Gaussians or splines \cite{hernandez1997,alfe2004a})
with a transformation of the single-particle orbitals into
localized Wannier-like functions can
achieve nearly linear scaling of the computational effort with the
system size when applied to insulators
\cite{williamson2001,alfe2004c,reboredo2005}.

\subsection{Antisymmetric forms with pair correlations}
\label{sec:wf_pfaff}

Apart from the Pauli exclusion principle, the Slater determinant
does not incorporate any correlations among the electrons, since it
is just an antisymmetrized form of a completely factorized
function, that is, of a product of one-body orbitals. A better account for
correlations can be 
achieved by wave functions built as the appropriate antisymmetrization
of a product of two-body orbitals. The resulting
antisymmetric forms are called pfaffians and can generally be
written as \cite{bajdich2008,bajdich2009}
% LM indexing
% \begin{equation}
% \label{eq:wf_pfaff}
% \Psi_{\rm A}^{\rm Pfaff}(\mathcal{R})=\mathcal{A}\Biggl[
% \prod_{m=1}^{N/2} \phi_m^{\alpha_m\beta_m}
%   \bigl(\bi{r}_{i}^{\sigma_i},\bi{r}_{j}^{\sigma_j}\bigr)
% \prod_{n=1}^{N-2N_{\rm P}}\psi_n^{\alpha_n}\bigl(\bi{r}_{k}^{\sigma_{k}}\bigr)
% \Biggr]\,,
% \end{equation}
%
% JK indexing
\begin{multline}
\label{eq:wf_pfaff}
\Psi_{\rm A}^{\rm Pfaff}(\mathcal{R})=\mathcal{A}\Biggl[
\prod_{m=1}^{N_{\rm P}} \phi_m^{\sigma_{m,1}\sigma_{m,2}}
  \bigl(\bi{r}_{m,1}^{\sigma_{m,1}},\bi{r}_{m,2}^{\sigma_{m,2}}\bigr)
\\ \times\kern -.5em
\prod_{n=1}^{N-2N_{\rm P}}\psi_n^{\sigma_n}\bigl(\bi{r}_{n}^{\sigma_{n}}\bigr)
\Biggr]\,,
\end{multline}
where $N_{\rm P}$ is the number of correlated pairs, $N_{\rm P}\leq N/2$.
The two-body orbitals $\phi_m^{\uparrow\downarrow}$ that couple
unlike-spin electrons (singlet pairs) are symmetric, whereas
$\phi_m^{\uparrow\uparrow}$ and $\phi_m^{\downarrow\downarrow}$
(triplet pairs) are antisymmetric functions. The inclusion of the one-body
orbitals $\psi_{n}^{\sigma}$ allows for
odd~$N$ or for only partially paired electrons. The pfaffian wave
functions can be viewed as compacted forms of particular multi-determinantal
expansions \eref{eq:wf_multidet}.

An important representative of the functional form \eref{eq:wf_pfaff}
is the Bardeen--Cooper--Schrieffer (BCS) wave 
function \cite{bardeen1957} projected onto a fixed number of particles,
which is obtained from~\eref{eq:wf_pfaff} by
considering a singlet pairing in an unpolarized system
($N_{\uparrow}=N_{\downarrow}=N/2$) with all two-body orbitals
identical. In that case, the antisymmetrization reduces to a
determinant \cite{bouchaud1988}
\begin{equation}
\label{eq:wf_bcs}
\Psi_{\rm A}^{\rm BCS}(\mathcal{R})=
\det\bigl[\phi^{\uparrow\downarrow}
  (\bi{r}_i^{\uparrow},\bi{r}_j^{\downarrow})\bigr]\,,
\end{equation}
where $\phi^{\uparrow\downarrow}
(\bi{r}_i^{\uparrow},\bi{r}_j^{\downarrow})$ is to be understood as a
$N/2\times N/2$ matrix $A_{ij}$. In the quantum-chemical literature,
this functional form is also known as the antisymmetrized geminal
power. Note that the form of the BCS wave function does not by itself imply
formation of Cooper pairs and their condensation, since the
determinant in \eref{eq:wf_bcs} reduces to the Slater wave function
\eref{eq:wf_slater} when the pair orbital is taken in the form
\begin{equation}
\label{eq:pairing_slater}
\phi_{\rm Slater}^{\uparrow\downarrow}
  (\bi{r}_i^{\uparrow},\bi{r}_j^{\downarrow})
=\sum_{n=1}^{N/2}\psi_n^{\uparrow}(\bi{r}_i^{\uparrow})
  \psi_n^{\downarrow}(\bi{r}_j^{\downarrow})\,.
\end{equation}
The BCS--Jastrow wave functions were employed in
investigations of ultra cold atomic gases (section~\ref{sec:app_bcs_bec})
\cite{carlson2003,giorgini2008} as well
as in calculations of the electronic structure of atoms \cite{casula2003}
and molecules \cite{casula2004}.

Trial wave functions with triplet pairing among particles were
suggested in the context of liquid ${}^3$He already two decades ago
\cite{bouchaud1987,bouchaud1988}. It was realized only recently
that even in these cases the exponentially large number of terms
constituting the pfaffian can be rearranged in a way that
facilitates its evaluation in a polynomial time, and therefore allows
application of the pfaffian--Jastrow trial wave functions in conjunction
with the VMC and DMC methods \cite{bajdich2006,bajdich2008}.

\subsection{Backflow coordinates}
\label{sec:wf_backflow}

Another way to further increase the variational freedom of the
antisymmetric part of the trial wave function is the
backflow transformation $\Psi_{\rm A}(\mathcal{R})\to \Psi_{\rm
  A}(\mathcal{X})$, where the new collective coordinates
$\mathcal{X}$ are functions of the original electron positions
$\mathcal{R}$. The designation ``backflow'' originates from an intuitive
picture of the correlated motion of particles introduced by Feynman to
describe excitations in quantum fluids \cite{feynman1954,feynman1956}.

In order to illustrate what is the origin of such coordinates, 
let us consider homogeneous interacting fermions in a periodic box 
with a trial wave function of the Slater--Jastrow type, 
$\Psi_{\rm T}(\mathcal{R})=\det\bigl[\exp(\rmi\bi{k}\cdot\bi{r}_i)\bigr]
\exp\bigl[\sum_{i<j}\gamma(r_{ij})\bigr]$.
The Jastrow factor is optimized so that its laplacian cancels out the
interactions 
as much as possible within the variational freedom.
Applying the kinetic energy operator to the Slater--Jastrow product
results in local energy of the form 
\begin{align}
\label{eq:backflowcoor}
&\frac{[\hat H \Psi_{\rm T}](\mathcal{R})}{\Psi_{\rm T}(\mathcal{R})}
= E_{\rm var}(\mathcal{R}) \nonumber\\
&\hskip1em -
\Bigl(\nabla\ln \bigl|\det\bigl[\exp(\rmi\bi{k}\cdot\bi{r}_i)\bigr]\bigr|\Bigr)
\!\cdot\! \biggl(\nabla\sum_{i<j}\gamma(r_{ij})\biggr)\,,
\end{align}
where we can qualitatively characterize $E_{\rm var}(\mathcal{R})$ as a mildly
varying function close to a constant while the second term represents 
a non-constant ``spurious'' contribution, which appears as a scalar
product of two fluxes.
Consider now the following modification of the Slater--Jastrow form,
$\Psi_{\rm T}(\mathcal{R})=\det\bigl[\exp(\rmi\bi{k}\cdot\bi{x}_i)\bigr]
\exp\bigr[\sum_{i<j}\gamma(r_{ij})\bigr]$,
where the single-particle coordinates are modified as 
$\bi{x}_i=\bi{r}_i +\sum_j\boldsymbol{\vartheta}(r_{ij})$ with
$\boldsymbol{\vartheta}(0)=\boldsymbol{0}$.
One can show that 
with a proper choice of the function
$\boldsymbol{\vartheta}(r_{ij})$, the laplacian of $\det\bigl[\exp(\rmi\bi{k}\cdot\bi{x}_i)\bigr]$ 
produces terms that cancel out most of the spurious contributions in the local 
energy given by \eref{eq:backflowcoor}. Of course, the backflow form generates
also new non-constant terms so that the wave function has
to be optimized for the overall maximum gain 
 using variational strategies.

In general, the new coordinates are written as
$\bi{x}_i=\bi{r}_i+\boldsymbol{\xi}_i(\mathcal{R})$
with $\boldsymbol{\xi}$ taken in a form analogous to the
parametrization of the Jastrow factor $U_{\rm corr}$
\eref{eq:wf_jastrow} and~\eref{eq:wf_jastrow_2body}. The vector
$\boldsymbol{\xi}$ contains
two-particle and possibly higher order correlations and, in systems
with external potentials, also inhomogeneous one-body terms. The backflow
transformation has been very successful in simulations and
understanding of homogeneous quantum liquids
\cite{lee1981,schmidt1981,kwon1998,holzmann2006}, and some progress has
recently been reported in applying these techniques also to atoms and
molecules \cite{lopezrios2006,bajdich2008}.

\section{Applications}
\label{sec:applications}

In the last part of the article we go through selected applications of
the quantum Monte Carlo methodology to the
electronic structure of solids. In practically all listed cases, with
the exception of sections~\ref{sec:heg} and~\ref{sec:app_bcs_bec}
dealing with model calculations, the
Slater--Jastrow functional form is employed as the trial wave function. The
reviewed results therefore
map out the accuracy that is achievable in the realistic solids when
the mean-field topology of the fermionic nodes is assumed. It is shown
that the quality of the DMC predictions is remarkable despite this
relatively simple approximation.

\subsection{Properties of the homogeneous electron gas}
\label{sec:heg}

\begin{SCtable*}
% Ecoh = (dH_solid-dH_gas) [kJ/mol] * 1000 /e/NA
%            e = 1.602e-19
%           NA = 6.022e23
\caption{\label{tab:cohesion}Cohesive energies of solids (in eV). Shown
  are DMC numbers unless only VMC data are available for the particular
  compound; those instances are 
  marked with~$(*)$. The latest
  results are preferred in cases where multiple calculations exist. If
  not indicated otherwise, the experimental 
  cohesive energies are deduced from the room-temperature formation
  enthalpies quoted in \cite{crc}.}
{\small\begin{tabular}{lr@{$\,$}c@{$\,$}l@{\ }lr@{\ }l}
\br
compound & \multicolumn{3}{c}{QMC} && \multicolumn{2}{c}{experiment} \\
\mr
% METALS
Li
 &  $1.09$&$\pm$&$0.05$ &\cite{sugiyama1989}
 &  1.65\\ %\cite{crc}  \\ \ms
 &  $1.57$&$\pm$&$0.01$ & \cite{yao1996} $(*)$\\ \ms
Na
 &  $1.14$&$\pm$&$0.01$ &\cite{kwee2008}
 &  $1.11$ \\ %\cite{crc} \\
 &  $1.0221$&$\pm$&$0.0003$ &\cite{maezono2003} & \\ \ms
Mg
 &  $1.51$&$\pm$&$0.01$ &\cite{pozzo2008a}
 &  $1.52$ \\ \ms %\cite{crc} \\ \ms
Al
 &  $3.23$&$\pm$&$0.08$ & \cite{gaudoin2002} $(*)$
 &  $3.43$ &\cite{gaudoin2002} \\ % extrapolated to 0 K, removed zero-point motion
\mr
% INSULATORS AND SEMICONDUCTORS
MgH${}_2$
 &  $6.84$&$\pm$&$0.01$ &\cite{pozzo2008a}
 &  6.83 \\ \ms %\cite{barin1995} \\ \ms (Barin and CRC give the same)
BN
 &  $12.85$&$\pm$&$0.09$ & \cite{malatesta1997} $(*)$  % 0 K
 &  $12.9\0$ &\cite{knittle1989} \\ \ms        % 300 K (\approx value)
C (diamond)
 &  $7.346$&$\pm$&$0.006$ &\cite{hood2003} % only ZPE in the DMC data
 &  7.37 &\cite{yin1981} \\ \ms % corrected to 0 temperature
% &&  7.409 \\ %\cite{crc}\\    % Barin gives the very same value (298.15K)
% Hood et al. (2003) provide $7.371\pm 0.005$ and reference
% JANAF Thermodynamic Tables, J. Phys. Chem. Ref. Data
% Vol. 14, Suppl. 1, edited by M.W. Chase, Jr. et al.
% (American Chemical Society, Washington, DC, 1985),
% pp. 61-65.
% I could not find any diamond data there (only graphite), the pages
% are bogus (aluminum data)
%
Si
 &  $4.62$&$\pm$&$0.01$ &\cite{alfe2004b}
 &  $4.62$ &\cite{farid1991} \\ % here we know error: $4.62\pm 0.08$
% &  $4.63$&$\pm$&$0.02$ &\cite{leung1999} & \\
 \ms
% &  $4.51\pm 0.03$ \cite{li1991} & \\ \ms % pseudohamiltonian (position dependent masses)
%
Ge  
 &  $3.85$&$\pm$&$0.10$ &\cite{rajagopal1995}
      % LDA FSE corrections+limited scaling, they don't give clear estimate of
      % errorbar
 &  3.86\\ \ms  %\cite{crc}  \\ \ms
GaAs
 &  $4.9$&$\pm$&$0.2$ &\cite{eckstein1996} $(*)$
 &  6.7\0 &\cite{arthur1967} \\  % crc gives this too
\mr
% TRANSITION METAL COMPOUNDS
MnO
 &  $9.29$&$\pm$&$0.04$ &\cite{kolorenc2010}
 &  $9.5\0$ \\ \ms %\cite{crc} \\
FeO
 &  $9.66$&$\pm$&$0.04$ &\cite{kolorenc2008}
 &  $9.7\0$ \\ \ms %\cite{crc} \\ \ms
NiO
 &  $9.442$&$\pm$&$0.002$ &\cite{needs2003} % quite lousy on FSE (16 atom
                                % cell, Gamma only)
 &  9.5\0 \\ \ms %\cite{crc} \\ \ms
BaTiO${}_3$
 &  $31.2$&$\pm$&$0.3$ &\cite{wagner2007} % FSE treatment unknown (very
                                % likely lousy)
 &  31.57 \\ %\cite{crc} \\ \ms
\br
\end{tabular}}
\end{SCtable*}

The homogeneous electron gas, also referred to as jellium, is one of
the simplest many-body models that can describe certain properties of
real solids, especially the alkali metals. At zero temperature, the
model is characterized by 
the densities of spin-up and spin-down electrons, $\rho_{\uparrow}$
and $\rho_{\downarrow}$, or, alternatively, by the total density
$\rho=\rho_{\uparrow}+\rho_{\downarrow}$ and the spin polarization
$\zeta=|\rho_{\uparrow}-\rho_{\downarrow}|/\rho$. 
It is convenient to express the
density~$\rho$ and other quantities in terms of a dimensionless parameter $r_{\rm
s}=[3/(4\pi\rho)]^{1/3}/a_{\rm B}$, where $a_{\rm B}$ is the Bohr radius. For
example, the density of valence electrons in the sodium metal
corresponds to $r_{\rm s}\approx 4$.

The total energy of jellium is particularly simple since it 
includes only
the kinetic energy of the
electrons, the Coulomb electron-electron repulsion, and a constant
which represents the 
interaction of the electrons with an inert uniformly distributed
positive charge that maintains overall charge neutrality of the
system. A straightforward
dimensional analysis shows that the kinetic energy
dominates the Coulomb interaction at high densities (small $r_{\rm
  s}$), where the electrons behave like a nearly ideal gas and the
unpolarized state ($\zeta=0$) is the most stable. In the limit of very
low densities, on the other  hand, the kinetic energy becomes
negligible and the electrons ``freeze'' into a Wigner crystal
\cite{wigner1934b}.

The homogeneous electron gas at zero temperature was one of the 
first applications of the variational and
diffusion Monte Carlo methods. In the early investigations
\cite{ceperley1978,ceperley1980}, only the unpolarized ($\zeta=0$) and
fully polarized ($\zeta=1$) fluid phases were considered
together with the Wigner crystal. Later, fluids with partial spin
polarization were taken into account as well
\cite{alder1982,ortiz1994,ortiz1999,zong2002}. The most accurate
trial wave functions (the Slater--Jastrow form with backflow
correlations) were used in reference \cite{zong2002} where it was
found that the unpolarized fluid is stable below $r_{\rm s}=50\pm
2$. At this density the gas undergoes a second-order phase transition
into a partially polarized state, and the spin
polarization~$\zeta$ then monotonically increases as the fluid is
further diluted. Eventually, the Wigner crystallization density is
reached, for which two DMC estimates exist: $r_{\rm s}=100\pm 20$
\cite{ceperley1980} and $r_{\rm s}=65 \pm 10$ \cite{ortiz1999}. The
discrepancy is presumably caused by the very small
energy differences between the competing phases over a wide range
of densities, and by uncertainties in the extrapolation to the
thermodynamic limit. 
Advanced finite-size extrapolation methods, outlined in
section~\ref{sec:pbc} earlier, could
possibly shed some new light on these quantitative differences.  
Indeed, recent calculations show further improvements in accuracy
of the total and correlation energies \cite{gurtubay2010}.
A number of static properties of the liquid phases that 
provide a valuable insight into the details of the electron
correlations in the jellium model and in Coulomb systems in general
were evaluated by QMC methods as well \cite{ortiz1994,zong2002,huotari2010,
gaudoin2010}.

The impact of the QMC calculations of the homogeneous electron 
gas \cite{ceperley1980} has been very significant because of the 
prominent position of the model as one of the simplest periodic
many-body systems, and also
due to the fact that the QMC correlation energy has become 
widely used as an input in density-functional calculations
\cite{perdew1981,perdew1992}.

The results quoted so far referred to the homogeneous Coulomb gas in
three dimensions. The two-dimensional gas, which is realized by
confining electrons to a surface, interface or to a thin layer in a
semiconductor heterostructure, has received similar if not even
greater attention of QMC practitioners
\cite{ceperley1978,tanatar1989,varsano2001,attaccalite2002,drummond2009a,drummond2009b,holzmann2009}.
The Wigner crystallization was predicted to occur at $r_{\rm s}=37\pm
5$ \cite{tanatar1989},%
\footnote{Note that in two dimensions
 the dimensionless density parameter $r_{\rm s}$ is
  defined as $r_{\rm s}=1/(a_{\rm B}\sqrt{\pi\rho})$.}
a value that coincides with the density $r_{\rm s}=35\pm 1$ where a
metal--insulator transition was experimentally observed later
\cite{yoon1999}.

\subsection{Cohesive energies of solids}
\label{sec:app_cohesion}

The cohesive energy measures the strength of the chemical
bonds holding the crystal together. It is defined as the difference
between the energy of a dilute gas of the constituent atoms or
molecules and the energy of the solid. Calculation of the
cohesive energy is a stringent test of the theory, since it has to
accurately describe two different systems with very dissimilar
electronic structure.

The first real solids whose cohesive energies were evaluated by a QMC
method were carbon and silicon in the diamond crystal structure
\cite{fahy1988,fahy1990}. These early VMC estimates were later refined
with the DMC method \cite{li1991,leung1999,alfe2004b,hood2003}. The
most accurate results to date are shown in table~\ref{tab:cohesion},
where we have compiled the cohesive energies of a number of compounds
investigated with the quantum Monte Carlo methods. Corresponding
experimental data are shown for comparison. The electronic total
energy calculated in QMC simulations is not the only contribution to
the cohesive energy of a crystal, and the zero-point and thermal motion of the
nuclei has to be accounted for as well, especially in compounds
containing light atoms. We refer the reader to the original references
for details of these corrections. At present, a
direct QMC determination of the phonon spectrum is generally not
practicable due to unresolved issues with reliable and efficient
calculation of forces acting on the nuclei \cite{badinski2010}. The effects due to the nuclear motion are thus typically estimated
within the density-functional theory.

Overall, the agreement of the DMC results with experiments is
excellent; the errors are smaller than $0.1$ eV most of the time,
including the Na and Mg elemental metals where coping with the
finite-size effects is more difficult. Notably, the diffusion Monte
Carlo performs (almost) equally well
 in strongly correlated
solids represented in table~\ref{tab:cohesion} by $3d$ transition
metal oxides MnO, FeO and NiO. The GaAs result is an obvious outlier
with a systematic error of almost 2 eV that the authors identify with
the deficiencies of their pseudopotentials \cite{eckstein1996}. The
two decades old application of the DMC method
to the Li metal \cite{sugiyama1989} is the only all-electron
simulation in the list and its comparison to a subsequent
pseudopotential calculation \cite{yao1996} suggests that a large part
of the discrepancy with the experiment is due to the fixed-node errors in the
high-density core regions. It is likely that a substantial
improvement would be observed if the all-electron calculations were
revisited with today's state of the art trial wave functions.

\subsection{Equations of state}
\label{sec:app_eos}

\begin{SCtable*}[1.0][t]
%
% 1 bar  = 100 kPa = 1e5  Pa
% 1 kbar = 1e3 bar = 1e8  Pa = 0.1 GPa
% 1 Mbar = 1e6 bar = 1e11 Pa = 100 GPa
%
% B  =  (C11  +  2C12)/3
%
% Ecoh = (dH_solid-dH_gas) [kJ/mol] * 1000 /e/NA
%            e = 1.602e-19
%           NA = 6.022e23
%
% Zincblende (B3) Structure is a two-component analog of the diamond (A4)
% structure, without the inversion symmetry in the middle of the bond
%
% 1 bohr = 0.529177249
%
\caption{\label{tab:eos_data}Equilibrium lattice constants $a_0$,
  equilibrium volumes~$V_0$ (per formula unit) and bulk moduli~$B_0$
  for a number of solids investigated with the QMC methods. The first
  line for each compound contains QMC predictions, the second line shows
  experimental data. Theoretical results for Li, Al and GaAs come from VMC
  simulations, the rest of the table corresponds to the DMC method.}
{\small\begin{tabular}{ll@{\ }ll@{\ }ll@{\ }l}
\br
compound & \multicolumn{1}{c}{$a_0$ (\AA)}&&
           \multicolumn{1}{c}{$V_0$ (\AA${}^3$)} &&
           \multicolumn{1}{c}{$B_0$ (GPa)} \\
\mr
% VMC
Li
  %& VMC \cite{yao1996}
  & $3.556\pm 0.005$ &\cite{yao1996}&&
  & $13\pm 2$ &\cite{yao1996} \\
  %& exp
  & 3.477 &\cite{berliner1986} &&
  & 12.8 &\cite{felice1977} \\ \bs
Al
  %& VMC \cite{gaudoin2002}
  & $3.970\pm 0.014$ &\cite{gaudoin2002}&&
  & $65\pm 17$ &\cite{gaudoin2002}\\
  %& exp \cite{gaudoin2002} %  extrapolated to 0 K, removed zero-point motion
  & 4.022 &\cite{gaudoin2002} &&
  & 81.3 &\cite{gaudoin2002} \\ \bs
GaAs 
  %& VMC \cite{eckstein1996}
  & $5.66\pm 0.05$ &\cite{eckstein1996}&&
  & $79\pm 10$ &\cite{eckstein1996} \\
  %& exp
  & 5.642 &\cite{predel1998} &&
  & $77\pm 1$ &\cite{cottam1973} \\
\mr
%
% DMC
LiH
  %& DMC \cite{binnie2009}
  & 4.006 &\cite{binnie2009} &&
  & $35.7\pm 0.1$ &\cite{binnie2009} \\
  %& exp \cite{loubeyre1998}
  & $4.07\pm 0.01$ &\cite{loubeyre1998} &&
  & $32.2\pm 0.03$ &\cite{loubeyre1998} \\ \bs
BN
%  & VMC \cite{malatesta1997} % 0 K
%  & $3.58\pm 0.04$
%  & $540\pm 150$ \\
%  & exp \cite{knittle1989}
%  & $3.615\pm 0.002$
%  & $369\pm 14$ \\ \bs
  %& DMC \cite{esler2010}
  &&& $11.812\pm 0.008$ &\cite{esler2010} %($V_0$)
  & $378\pm 3$ &\cite{esler2010} \\
  %& exp \cite{datchi2007}
  &&& $11.812\pm 0.001$ &\cite{datchi2007}
  & $395 \pm 2$ &\cite{datchi2007} \\ \bs
Mg
  %& DMC \cite{pozzo2008a}
  &&& $23.61\pm 0.04$ &\cite{pozzo2008a} %$(V_0)$
  & $31.2\pm 2.4$ &\cite{pozzo2008a} \\
  %& exp
  &&& 23.24 &\cite{walker1959} % [{\small www.webelements.com}]
  & $36.8\pm 3.0$ &\cite{errandonea2003} \\ \bs
MgO
  %& DMC \cite{alfe2005} % includes lattice vibrations and pp error estimate
  & 4.23 &\cite{alfe2005} &&
  & 158 &\cite{alfe2005} \\
  %& exp \cite{fei1999}
  & 4.213 &\cite{fei1999} &&
  & $160\pm 2$ &\cite{fei1999} \\ \bs
MgH${}_2$
  %& DMC \cite{pozzo2008a}
  &&& $30.58\pm 0.06$  &\cite{pozzo2008a} %$(V_0)$
  & $39.5\pm 1.7$ &\cite{pozzo2008a} \\
  %& exp
  &&& 30.49 &\cite{bortz1999}
  & --- \\ \bs
%
%Ne
%  & DMC \cite{drummond2006}
%  & 19.046 ($V_0$)
%  & 2.754 \\
%  & exp \cite{hemley1989}
%  & 22.241
%  & 1.097 \\ \bs
%
C
  %& DMC \cite{maezono2007}
  & $3.575\pm 0.002$ &\cite{maezono2007} && % temp. and ZPE in the DMC data
  & $437\pm 3$ &\cite{maezono2007} \\     % temp. and ZPE in the DMC data
(diamond)  %& exp
  & 3.567 &\cite{crc} &&
  & $442\pm 4$ &\cite{mcskimin1972} \\ \bs
Si
  %& DMC \cite{alfe2004b}
  & $5.439\pm 0.003$ &\cite{alfe2004b} &&
  & $103\pm 10$ &\cite{alfe2004b} \\
%  & DMC~\cite{li1991} % pseudohamiltonian (position dependent masses)
%  & $5.45\pm 0.02$ &
%  & $103\pm 7$ \\
  %& exp
  & 5.430 &\cite{okada1984} && % extrapolated to 0 K
  & {\0}99.2 &\cite{hall1967} \\ \bs  % extrapolated to 0 K
SiO${}_2$
  &&& $37.6\pm 0.3$ &\cite{driver2010}
  & $32\pm6$ &\cite{driver2010}\\
(quartz)
  &&& $37.69$ &\cite{hanzen1989}
  & $34$ &\cite{hanzen1989} \\ \bs
FeO
  %& DMC \cite{kolorenc2008}
  & $4.324\pm 0.006$ &\cite{kolorenc2008} &&
  & $170 \pm 10$ &\cite{kolorenc2008} \\
  %& exp
  & 4.334 &\cite{mccammon1984} &&
  & $\approx 180$ &\cite{zhang2000} \\
\br
\end{tabular}}
\end{SCtable*}

\begin{SCfigure*}
\resizebox{1.2\columnwidth}{!}{%
\includegraphics{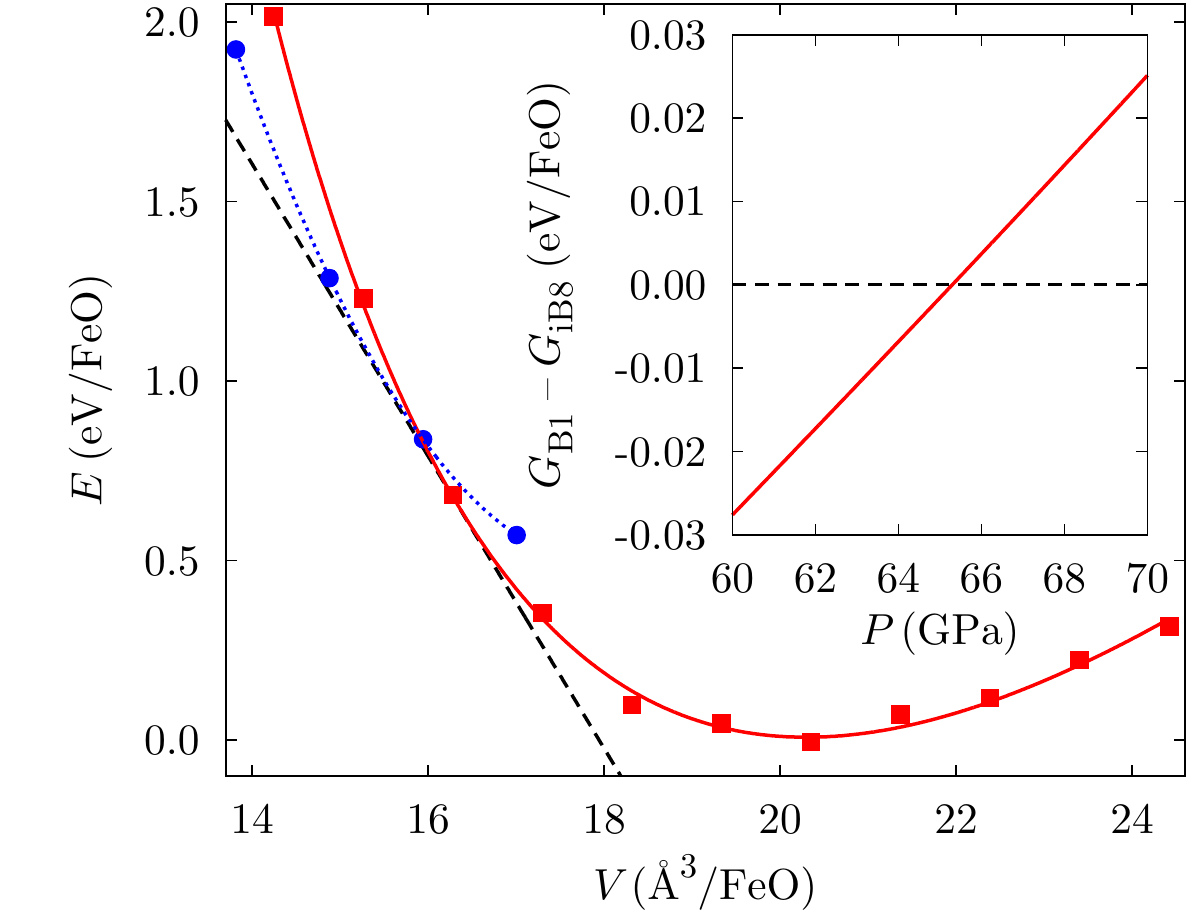}}
\caption{\label{fig:feo_eos} DMC total energies of
  the rock-salt (squares) and the NiAs (circles) phases of
  FeO. Statistical error bars are smaller than the symbol sizes. Lines
  are fits with the Murnaghan equation of state. Inset: Difference between
  the Gibbs potentials of the two phases at $T = 0$ K; its intercept with
  the $x$ axis determines the transition pressure $P_{\rm t}$. Adopted from
  \cite{kolorenc2008}.}
\end{SCfigure*}

The equilibrium volume $V_0$, the lattice constant $a_0$, and the bulk
modulus $B_0=V(\partial E/\partial V)|_{V=V_0}$ are among the most
basic parameters characterizing elastic properties of a solid near the
ambient conditions. Within QMC methods, these quantities are determined
by evaluating the total energy at several volumes around~$V_0$ and
by fitting an appropriate model \cite{murnaghan1944,vinet1986} of the
equation of state $E(V)$ through the acquired data, see
figure~\ref{fig:feo_eos} for an illustration. Results of 
this procedure for a wide range of solids are shown in
table~\ref{tab:eos_data} together with the corresponding experimental
data. As in the case of the cohesion energy discussed in the preceding
section, the raw QMC numbers correspond to the static lattice and
corrections due to the motion of nuclei may be needed to facilitate
the comparison with experimentally measured quantities. Particular
details about applied adjustments can be inspected in the original
papers.

The data in table~\ref{tab:eos_data} demonstrate that the equilibrium
geometries predicted by the QMC simulations are very good and all lie
within 2 \% from the experiments, in many cases within only a few
tenths of a percent. The agreement is slightly worse for the bulk
moduli, where errors of several percent are common and in a few
instances the mean values of the Monte Carlo estimates deviate from
the experimental numbers by more than 10 \%. Note, however, that
determination of the curvature of $E(V)$ near its minimum is
impeded by the stochastic noise of the QMC energies and that the error
bars on the less favourable results are relatively large.

Quantum Monte Carlo methods are not limited to the covalent solids
listed in table~\ref{tab:eos_data}. Investigation of the equation of
state of solid neon \cite{drummond2006} represents an application to a
crystal bound by van der Waals forces. Although the shallowness of the
minimum of the energy--volume curve in combination with the Monte
Carlo noise did not allow to determine the lattice constant and the
bulk modulus to a sufficient accuracy, the DMC equation of state was
still substantially better than results obtained with LDA and
GGA. This example together with a recent study of interlayer binding
in graphite \cite{spanu2009} illustrate that the diffusion Monte Carlo
method provides a fair description of dispersive forces already with
the simple nodal structure defined by the single-determinantal
Slater--Jastrow wave function. More accurate trial wave functions
incorporating backflow correlations were employed to study van der Waals
interactions between idealized metallic sheets and wires
\cite{drummond2007}.

Calculations of the equations of state are by no means restricted to the
vicinity of the equilibrium volume, and many of the references quoted in
table~\ref{tab:eos_data} study the materials up to very high
pressures. Such investigations are stimulated by open problems
from Earth and planetary science as well as from other areas of materials
physics. Combination of the equation of state with the
pressure dependence of the Raman frequency
\cite{maezono2007,esler2010}, both calculated from first principles
with QMC, can provide a very accurate high-pressure
calibration scale for use in experimental studies of condensed matter
under extreme conditions \cite{esler2010}.

\subsection{Phase transitions}
\label{sec:app_transitions}

Theoretical understanding of structural phase transitions often
necessitates a highly accurate description of the involved crystalline
phases. Simple approximations are known to markedly 
fail in a number of instances due to significant changes in
the bonding conditions across the transition. A classical example is the
quartz--stishovite transition in silica (SiO${}_2$), where LDA performs
very poorly and GGA is needed to reconcile the DFT picture with
experimental findings \cite{hamann1996}. The diffusion Monte Carlo
method has been employed to investigate pressure-induced phase
transitions in Si \cite{alfe2004b}, MgO \cite{alfe2005}, FeO
\cite{kolorenc2008} and SiO${}_2$ \cite{driver2010}.

A transition from the diamond structure to the
$\beta$-tin phase in silicon was estimated to occur at $P_{\rm t}=16.5\pm
0.5$ GPa \cite{alfe2004b}, which lies outside the range of
experimentally determined values 10.3--12.5~GPa (see \cite{alfe2004b}
for compilation of experimental literature). Since the diamond
structure is described very accurately with the DMC method as
testified by the data in tables~\ref{tab:cohesion}
and~\ref{tab:eos_data}, it was suggested
that the discrepancy is a manifestation of the fixed-node errors in
the high-pressure $\beta$-tin phase. This view is supported by a
recent calculation utilizing the so-called phaseless auxiliary-field
QMC, a projector Monte Carlo method that
shows smaller biases related to the fermion sign problem in this
particular case and predicts the transition at $12.6\pm 0.3$ GPa
\cite{purwanto2009}. It should be noted, however, that the volume
 at which the transition occurs
 was fixed to its experimental value in
this later study, whereas the approach pursued in \cite{alfe2004b} was
entirely parameter-free.

In iron oxide (FeO), a transition from the rock-salt structure to
a NiAs-based phase is experimentally observed to occur around 70 GPa
at elevated temperatures \cite{jeanloz1980} and to move to higher
pressures exceeding 100 GPa when the temperature is lowered
\cite{fei1994}. The DMC simulations summarized in
figure~\ref{fig:feo_eos} place the transition at $P_{\rm t}=65\pm 5$
GPa \cite{kolorenc2008}. This value represents a significant
improvement over LDA and GGA that both find the NiAs structure more
stable than the rock-salt phase at all volumes. The agreement with
experiments is nevertheless not entirely satisfactory, since the DMC
prediction corresponds to zero temperature where experimental
observations suggest stabilization of the rock-salt structure
to higher pressures. Sizeable sensitivity of 
the transition pressure $P_{\rm t}$ to the choice of the one-particle
orbitals in the Slater--Jastrow trial wave function was demonstrated
in a subsequent study \cite{kolorenc2010}, but those wave functions
that provided higher $P_{\rm t}$ also increased the total energies, and
therefore represented poorer approximations of the electronic ground
state. It remains to be determined, whether the
discrepancy between the experiments and the DMC
simulations is due to inaccuracies of the Slater--Jastrow nodes or if
some physics not included in the investigation, for instance
the inherently defective nature of the real FeO crystals, plays a
significant role.

Investigations of phase transitions involving a liquid phase, such as
melting, are considerably more involved due to a non-trivial motion of
ions. An often pursued approach is a molecular dynamics simulation of
ions subject to forces derived from the electronic ground state
that is usually approximated within the density-functional
theory. More accurate results would be achieved if the forces were
calculated using quantum Monte Carlo methods instead. At present, this
is generally not feasible due to excessive noise of the available
force estimates \cite{badinski2010}. Nevertheless, it was demonstrated
that one can obtain an improved picture of the energetics of the
simulated system when its electronic energy is evaluated with the aid of a QMC
method while still following the ion trajectories provided by the DFT
forces \cite{grossman2005,sola2009b}.

\subsection{Lattice defects}
\label{sec:app_defects}

The energetics of point defects substantially influences high-temperature
properties of crystalline materials. Experimental investigations of
the involved processes are relatively difficult, and it would be
very helpful if the electronic structure theory could provide dependable
predictions. The role of electron correlations in point defects was
investigated with the DMC method in silicon
\cite{leung1999,batista2006} and in diamond \cite{hood2003}. The
formation energies of selected self-interstitials in silicon were found about
$1.5$ eV larger than in LDA, whereas the formation energy of vacancies
in diamond came out as approximately 1 eV smaller than in LDA. These
differences represent 20--30 \% of the formation energies and indicate
that an improved account of electron correlations is necessary for
accurate quantitative understanding of these phenomena.
% It was demonstrated that modern hybrid exchange-correlation
% functionals can come 
% close to the DMC predictions \cite{batista2006}, which can be
% understood as an indirect verification that the observed discrepancies
% between the DMC method and the earlier DFT treatments are indeed not due to
% fixed-node errors in the Monte Carlo simulations.

Charged vacancies constituting the Schottky defect were investigated
in MgO \cite{alfe2005b}, and in this case the predictions of the DMC
method differ only marginally from the results obtained within the
local density approximation. The non-zero net charge of the supercells
employed in these simulations represents an additional technical
challenge in the form of increased finite-size effects that require a
modification of some of the size extrapolation techniques discussed in
section~\ref{sec:pbc} \cite{leslie1985,makov1995}.

\subsection{Surface phenomena}

\begin{SCtable*}[50]
\caption{\label{tab:jelliumsurface}
Comparison of the surface energies (in erg cm${}^{-2}$) of the
homogeneous electron gas calculated by a number of electronic
structure methods \cite{wood2007}. The DMC calculations were done with
the LDA orbitals in the trial wave functions.}
{\small\begin{tabular}{lrrrr}
\br
$r_{\rm s}$ & LDA & GGA & \multicolumn{1}{c}{DMC} & RPA \\
\mr
2.07 & $-608.2$ & $-690.6$ & $-563 \pm 45$ & $-517$ \\
2.30 & $-104.0$ & $-164.1$ & $-82 \pm 27$  &  $-34$ \\
2.66 & \m 170.6 & \m 133.0 & $179 \pm 13$  &   216  \\
3.25 & \m 221.0 & \m 201.2 & $216 \pm 8\0$   &   248  \\
3.94 & \m 168.4 & \m 158.1 & $175 \pm 8\0$   &   182  \\
\br
\end{tabular}}
\end{SCtable*}

Materials surfaces are fascinating systems from the point of view
of electronic structure and correlation effects. 
The vacuum boundary condition provides surface atoms with more space to
relax their positions and
surface
electronic states enable the electronic structure to develop features 
which cannot form in the periodic bulk.   
This leads to a plethora 
of surface reconstruction possibilities with perhaps the most studied 
paradigmatic case of $7\times7$ Si(111) surface reconstruction.
Seemingly, QMC methods should be straightforward to apply to these systems, 
similarly to the
three-dimensional periodic solids. However, mainly 
technical reasons make such 
calculations quite difficult. 
There are basically two possibilities how to model 
a surface. One option is to use a two-dimensional
periodic slab geometry which requires certain minimal slab
thickness in order to accurately represent the bulk environment for
the surface layers on both sides.
The resulting simulation cells end up quite
large making many such simulations out of reach at present. 
The other option is to use a cluster with appropriate termination that
mimics the bonding pattern of the bulk atoms. This strategy assumes
that the termination does not affect the surface geometries in a
substantial manner. Moreover, it is applicable only to insulating
systems. Given these difficulties, the   
 QMC simulations of surfaces are rare and this research area
awaits to  be explored in future.

The simplest possible model for investigation of surface physics is
the surface of the homogeneous electron gas that has been studied
by DFT as well as QMC methods. 
 The first QMC calculations \cite{acioli1996} were later found 
to be biased due to complications arising from finite-size effects,
especially due to different scaling of finite-size corrections
for bulk and surface. Once these issues have been properly taken
into account by Wood and coworkers \cite{wood2007}, the QMC results
have exhibited  trends that were consistent with
DFT and RPA methods which are expected to perform 
reasonably well for this model system 
(see table~\ref{tab:jelliumsurface}).

Applications to real materials surfaces are  still very few. 
The cluster model was used in calculations of Si(001)  surface by
Healy {\it et al.} \cite{healy2001} with the goal of  elucidating  
a long-standing 
puzzle in reconstruction geometry of this surface, which exhibits 
regularly spaced rows of Si-Si dimers. The dimers could take two
possible conformations: They can be either positioned symmetrically or
form an alternating buckling in a zig-zag fashion.
While experiments suggested the buckled geometry as the low-temperature
ground state, theoretical
calculations produced conflicting results, in which various
methods favoured one or the other structure. 
The QMC calculations \cite{healy2001} concluded that the buckled geometry
is lower by about 0.2 eV/Si$_2$. This problem was studied with
QMC methodology also by 
Bokes and coworkers \cite{bokes2002} who found 
that several systematic errors (such as uncertainty of
geometries in cluster models and pseudopotential biases) added to about 
0.2 eV, and therefore prevented unequivocal determination of the most stable geometry. 
This conclusion corroborated the experimental findings which suggested that
at temperatures above 100 K the distinct features of buckling were largely washed out
and indicated that the effect is energetically very small.
Very recently, the QMC study of this system has been repeated by Jordan and coworkers
\cite{lampart2008} with
the conclusion that the buckled structure is lower by about 0.1 eV/Si$_2$ and that
the highest level correlated basis set method which they used
(CASPT3) is consistent with this finding.
It was also clear that once the correlations were taken into account,
the energy differences between the competing surface reconstruction patterns
were becoming very small. This brings the calculations closer to reality, where the two 
structures could be within a fraction of 0.1 eV/Si$_2$ as suggested by experiments.

Perhaps the most realistic QMC calculations of surfaces have been done on LiH and MgO surfaces
by the group of Alf\`e and Gillan \cite{alfe2006,binnie2009} who
compared predictions of several DFT functionals with the fixed-node DMC method.
The results showed significant differences between various
exchange-correlation functionals. 
For the MgO(100) surface the best agreement with QMC results was found for
the LDA functional, while for the LiH surface the closest agreement between QMC and DFT
predictions was found for particular GGA functionals.
% for PW91 and PBE functionals. 

Clearly, more applications are needed to assess the effectiveness of QMC approaches 
for investigation of surface physics. As we have already mentioned, the
surfaces represent quite challenging systems for QMC methods. Nevertheless, we expect
more applications to appear in the future
 since the field is very rich  
in variety of correlation effects that are difficult to capture by
more conventional methods.
   
\subsection{Excited states}

\begin{SCtable*}[50]
\caption{\label{tab:oxides_gap}Band gaps (in eV) of Mott insulators MnO and
  FeO calculated with the fixed-node DMC method. Experimental
  data are provided for comparison.}
{\small\begin{tabular}{lcr}
\br
compound & DMC & \multicolumn{1}{c}{experiment} \\
\mr
MnO & $4.8\pm 0.2$ \cite{mitas2010}
    & $3.9\pm 0.4$ \cite{vanelp1991}\\
FeO & $2.8\pm 0.3$ \cite{kolorenc2008}
    & $\approx 2.4$ \cite{bowen1975}\\
\br
\end{tabular}}
\end{SCtable*}

The VMC and the fixed-node DMC methods both build on the variational
principle, and they therefore seem to be applicable exclusively to the
ground-state properties. Nevertheless, the variational principle can
be symmetry constrained, in which case the algorithms search for the
lowest lying eigenstate within the given symmetry class (provided, in
the case of the DMC method, that the eigenstate is non-degenerate
\cite{foulkes1999}), and thus enable access to selected excited states.

Excitation energies in solids are calculated as differences between the
total energy obtained for the ground state and for the excited state.
It is a computationally demanding procedure
 since the stochastic
fluctuations of the total energies are proportional to the number of
electrons in the simulation cell, whereas the excitation energy is an
intensive quantity. Trial wave functions for excited states
are formed by modifying the determinantal part of the
ground-state Slater--Jastrow wave function such that an occupied
orbital in the ground-state determinant is replaced by a virtual
orbital. This substitution corresponds to an optical absorption
experiment where an 
electron is excited from the valence band into the conduction
band. The fact that both the original occupied orbital and the new
virtual orbital necessarily belong to the same $K$ point restricts the
types of excitations that can be 
studied, since only a limited number of $k$ points from the
primitive cell fold to the given $K$ point of the simulation cell,
recall equations \eref{eq:bloch_supercell}--\eref{eq:pack_monkhorst}. 
Clearly, the larger the
simulation cell, the finer mapping out of excitations can
be performed.

Averaging over twisted boundary conditions (section~\ref{sec:tabc}) is
not applicable to the calculations of the
excitation energies, since both the ground state and the excited state
are fixed to a single $K$ point. This is not a significant issue, since
finite-size effects tend to cancel very efficiently in the differences
of the total energies calculated at the same $K$ point.

DMC simulations following the outlined recipe were utilized to
estimate the band gap in solid nitrogen \cite{mitas1994} and in
transition-metal oxides FeO \cite{kolorenc2008} and MnO
\cite{mitas2010}. The gaps calculated for the two strongly correlated
oxides are 
compared with experimental data in table~\ref{tab:oxides_gap}. The
ratio of the FeO and MnO gaps is reproduced quite well, but the DMC gaps
themselves are somewhat overestimated, likely due to inaccuracies of
the trial wave functions used for the excited states. A large number
of excitations were calculated in silicon \cite{williamson1998} and in
diamond \cite{mitas1996,towler2000}, and the obtained data were used
to map, albeit sparsely, the entire band structure. In these weakly
correlated solids the agreement with experiments is very
good. Recently, a pressure-induced insulator--metal transition was
investigated in solid helium by calculating the evolution of the band
gap with compression \cite{khairallah2008}. As illustrated in
figure~\ref{fig:helium_gap} ({\itshape copyrighted material; not
  available in this version; see figure~1 in
  \cite{khairallah2008}\/}), the DMC band gaps were found to
practically coincide with the gaps calculated with the $GW$ method.

\addtocounter{figure}{1}
\makeatletter
\protected@write\@auxout{}%
         {\string\newlabel{fig:helium_gap}{{\thefigure}{}{}{}{}{}}}%
\makeatother

% \begin{figure*}
% \resizebox{.75\textwidth}{!}{%
% \includegraphics{figs/helium_gap}}
% \caption{\label{fig:helium_gap} Band gap in hcp solid helium as a
%   function of density calculated with several electronic structure
%   methods. The gap is determined as the energy needed to promote an
%   electron from the highest valence state at the 
%   $\Gamma$ point to the lowest conduction state at the $M$ point. The
%   DMC data from the smaller simulation cell (8 atoms)
%   show a significant finite-size error, the data from the larger cell
%   (64 atoms) are converged with respect to the system size and virtually
%   coincide with the $GW$ predictions. Metallization densities estimated
%   by the individual methods are indicated as well. Adopted with
%   permission from \cite{khairallah2008}.}
% \end{figure*}

\subsection{BCS--BEC crossover}
\label{sec:app_bcs_bec}

\begin{SCfigure*}
\resizebox{1.2\columnwidth}{!}{%
\includegraphics{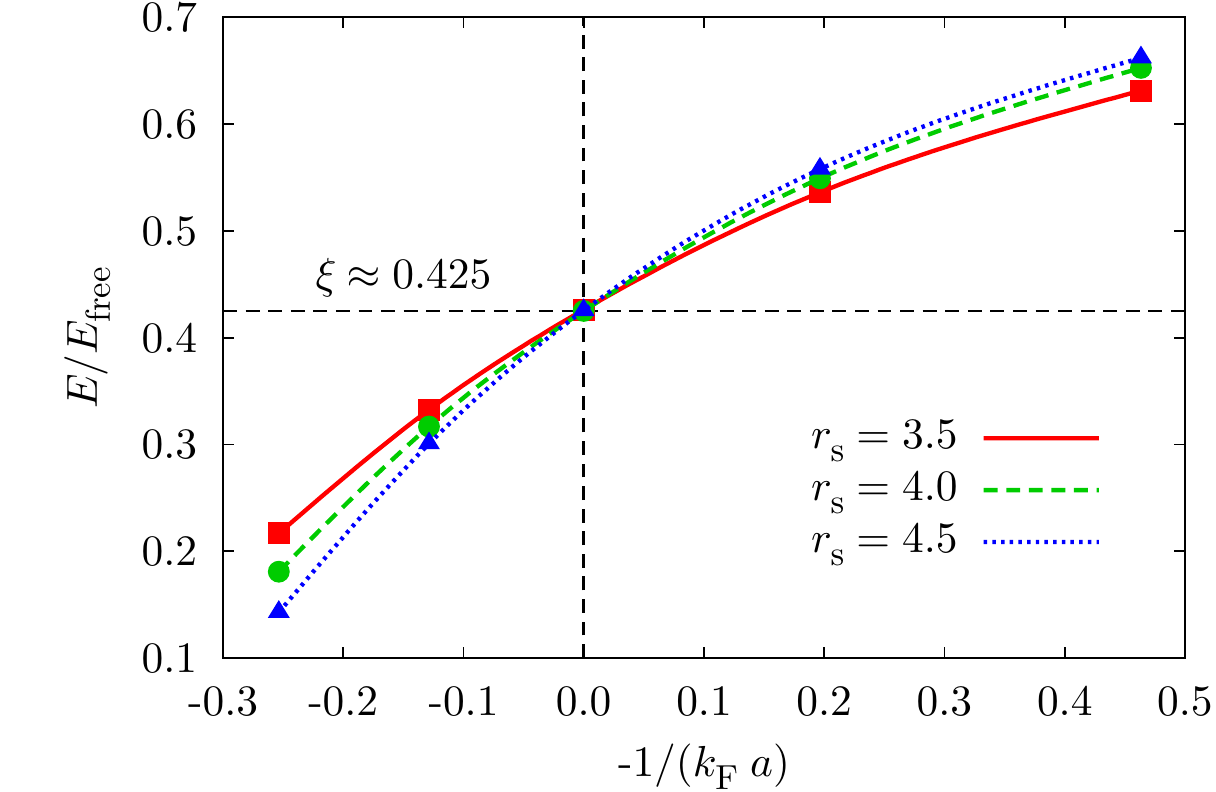}}
\caption{\label{fig:bcs_bec_energy} Fixed-node DMC energies of 38
  fermions in a cubic box with the periodic boundary
  conditions plotted as a function of interaction strength. BEC regime
  is on the left, BCS limit on the right. Shown 
  are three particle densities~$\rho$ characterized 
  by the dimensionless parameter~$r_{\rm s}$ defined in
  section~\ref{sec:heg}. The simulations
  employed BCS--Jastrow trial wave function. Statistical error bars
  are smaller than the symbol sizes. Data taken from
  \cite{bajdich2010}.}
\end{SCfigure*}

The repulsive Coulomb interaction considered so far is not the only
source of non-trivial many-body effects in the electronic structure
of solids. A weak attractive interaction among electrons is
responsible for a very fundamental phenomenon---the electronic states
in the vicinity of the Fermi level rearrange into bosonic Cooper pairs
that condense and give rise to superconductivity. The ground state of
the system can be described by the BCS
wave function $\Psi_{\rm BCS}$ discussed earlier in
section~\ref{sec:wf_pfaff} \cite{bardeen1957}. The Cooper pair is an
entity that has a meaning only as a constituent of $\Psi_{\rm
BCS}$. In order to form an isolated two-electron bound state, some
minimal strength of the two-body potential is needed in three
dimensions, whereas the Cooper instability itself occurs for
arbitrarily weak attraction. When the interaction is very strong, the
composite bosons formed as the two-electron bound 
states indeed exist and undergo the Bose--Einstein condensation
(BEC). It turns out that a mean-field description of both the BCS and
BEC limits leads to the same form of the many-body wave function,
which indicates that the interacting fermionic system is likely to
continuously evolve from one limit to the other when the interaction
strength is gradually changed 
\cite{eagles1969,leggett1980,nozieres1985}. A large amount of research
activity aimed at detailed understanding of this physics was
stimulated by the possibility to realize the BCS--BEC crossover in
experiments with optically trapped ultra-cold atoms \cite{ketterle2008}.

In a dilute Fermi gas with short-ranged spherically symmetric inter-particle
potentials, the interactions are fully characterized by a single
parameter, the two-body scattering length $a$. The system is
interpolated from the BCS regime to the BEC limit by varying $1/a$ from
$-\infty$ to $\infty$. In experiments, this is achieved by tuning across
the Feshbach resonance with the aid of an external magnetic field. Particularly
intriguing is the quantum state of an unpolarized homogeneous gas at
the resonance itself, where the 
scattering length diverges ($1/a=0$). The only relevant length scale
remaining in the problem in this case is the inverse of the Fermi wave vector
$1/k_{\rm F}$, and all ground-state properties should therefore be universal 
functions of the Fermi energy $E_{\rm F}$. Since there is only a single
length scale, the system is said to be in the unitary limit. The total
energy can be written as
\begin{equation}
E=\xi E_{\rm free}=\xi\,\frac{3}{5} E_{\rm F}\,,
\end{equation}
where $E_{\rm free}$ denotes the energy of a non-interacting system
and $\xi$ is a universal parameter. The universality of $\xi$ is
illustrated in figure~\ref{fig:bcs_bec_energy} that shows the ratio
$E/E_{\rm free}$ as a function of the interaction strength calculated
for three different particle densities using the diffusion
Monte Carlo method with the trial wave function of the
BCS--Jastrow form. All three curves indeed intersect at $1/a=0$
with the parameter $\xi$ estimated as $0.42\pm 0.01$
\cite{carlson2003,astrakharchik2004,chang2004,morris2010}. The energy
calculated 
with the fermionic nodes fixed by the Slater--Jastrow wave function is
considerably higher and would lead to $\xi\approx 0.54$
\cite{carlson2003}, which underlines the significance of particle
pairing in this system.

A further insight into the formation of the Cooper pairs is provided
by evaluation of the condensate fraction that can be estimated
from the off-diagonal long-range order occurring in the two-particle
density matrix \cite{yang1962}. The condensate fraction $\alpha$ is
given as
\begin{equation}
\alpha=\frac{N}{2}\,\lim_{r\to\infty}
\rho^{\rm P}_{2\uparrow\downarrow}(r)
\end{equation}
and the so-called projected two-particle density matrix
$\rho^{\rm P}_{2\uparrow\downarrow}$ is \cite{depalo2002}
\begin{align}
\rho^{\rm P}_{2\uparrow\downarrow}(\bi{r})
=\frac1{4\pi}&\int \rmd\Omega_{\bi{r}}\int \rmd^{3N}\mathcal{R}
\nonumber \\
&\times \Psi^*(\bi{r}_1+\bi{r},\bi{r}_2+\bi{r},\bi{r}_3,\ldots,\bi{r}_N)
\nonumber \\
&\times \Psi(\bi{r}_1,\bi{r}_2,\bi{r}_3,\ldots,\bi{r}_N)\,,
\end{align}
where $\bi{r}_1$ corresponds to the spin-up state and $\bi{r}_2$ to
the spin-down state. The evolution of $\alpha$ with interaction
strength calculated with the DMC method \cite{astrakharchik2005} is
shown in figure~\ref{fig:condensate_fraction} ({\itshape copyrighted
material; not available in this version; see figure~4 in
\cite{astrakharchik2005}\/}). It is found that
approximately half of the particles participates in pairing in the
unitary regime and this fraction quickly decreases towards the BCS
limit, where only states in the immediate vicinity of the Fermi level
contribute to the Cooper pair formation. Note that the condensate
fraction vanishes if the Slater--Jastrow form is used in place of the
trial wave function.

\addtocounter{figure}{1}
\makeatletter
\protected@write\@auxout{}%
         {\string\newlabel{fig:condensate_fraction}{{\thefigure}{}{}{}{}{}}}%
\makeatother

% \begin{figure*}
% \resizebox{.75\textwidth}{!}{%
% \includegraphics{figs/condensate_fraction}}
% \caption{\label{fig:condensate_fraction} Condensate fraction in a
%   dilute Fermi gas as a function of the inter-particle interaction
%   strength specified by the two-body scattering length $a$
%   (interaction increases from right to left as in
%   figure~\ref{fig:bcs_bec_energy}). Compared are: fixed-node
%   DMC simulations (symbols) and a self-consistent mean-field theory
%   \cite{salasnich2005} (line). Adopted with permission from
%   \cite{astrakharchik2005}.}
% \end{figure*}

The diffusion Monte Carlo simulations were used to study also the
total energy and the 
particle density profile in the unitary Fermi gas subject to a harmonic
confining potential \cite{chang2007,blume2007,vonstecher2008}. Due to
the lowered symmetry  compared to the homogeneous calculations
referred above, the system sizes were more limited. To extrapolate the
findings to a larger number of particles, a density functional theory
fitted to the DMC data can be employed \cite{salasnich2008}.

\section{Concluding remarks}
\label{sec:conclusions}

In this article we have attempted to provide an overview of selected
quantum Monte Carlo methods that facilitate calculation of various
properties of correlated quantum systems to a very high
accuracy. Particular attention has been paid to technical details
pertaining to applications of the methodology to extended systems such
as bulk solids. We hope that we have been able to demonstrate that the
QMC methods, thanks to their accuracy and a wide range of applicability,
represent a powerful and valuable alternative to more traditional ab initio
computational tools.

\subsection*{Acknowledgments}

We thank G.~E. Astrakharchik and B. Militzer for
providing their data, and K.~M. Rasch for suggestions to the
manuscript. J.~K. would like to acknowledge financial support by the
Alexander von Humboldt foundation during preparation of the
article. Support of L.~M.  research by NSF EAR-05301110,
DMR-0804549 and OCI-0904794 grants and by DOD/ARO and DOE/LANL
DOE-DE-AC52-06NA25396 grants is gratefully acknowledged. We
acknowledge also allocations at ORNL through INCITE and CNMS
initiatives as well as allocations at NSF NCSA and TACC centers.

\end{multicols}

\hskip .25\textwidth\vrule height .5pt width .5\textwidth depth 0pt
\begin{multicols}{2}
\let\bibfont=\small           % Bibliography's font size
\setlength{\bibsep}{2.5pt}             % Separation between BiB entries 
\renewcommand\refname{References}
\bibliography{qmc_review}
\end{multicols}

\end{document}